\documentclass[12pt, draftclsnofoot, onecolumn]{IEEEtran}

\usepackage{graphicx}
\usepackage[noadjust]{cite}
\usepackage{color}
\usepackage{amsmath,amssymb,mathtools}
\usepackage{subcaption}
\usepackage{fancyvrb, dsfont}

\IEEEoverridecommandlockouts

\newcommand{\upperRomannumeral}[1]{\uppercase\expandafter{\romannumeral#1}}

\begin{document}

\title{Wireless Channel Dynamics and Robustness for \\ Ultra-Reliable Low-Latency Communications}

    \author{
    \vspace{-10pt}
    \IEEEauthorblockN{Vasuki~Narasimha~Swamy\IEEEauthorrefmark{1}, Paul~Rigge\IEEEauthorrefmark{1}, Gireeja~Ranade\IEEEauthorrefmark{1}\IEEEauthorrefmark{2}, \\ Borivoje~Nikoli\'{c}\IEEEauthorrefmark{1}, Anant~Sahai\IEEEauthorrefmark{1} }\\
    \IEEEauthorblockA{\IEEEauthorrefmark{1}University of California, Berkeley, CA, USA, \\ \IEEEauthorrefmark{2}Microsoft Research, Redmond, WA, USA}
}
\maketitle
\vspace{-40pt}
\begin{abstract}
\vspace{-10pt}
Interactive, immersive and critical applications demand ultra-reliable
low-latency communication (URLLC).
To build wireless communication systems that can support these
applications, understanding the characteristics of the wireless medium is paramount.
Although wireless channel characteristics and dynamics have been
extensively studied, it is important to revisit these concepts in the
context of the strict demands of low latency and ultra-reliability.
In this paper, we bring a modeling approach from robust control to
wireless communication --- the wireless channel characteristics are given a
nominal model around which we allow for some quantified
uncertainty. We propose certain key ``directions'' along which to
bound model uncertainty that are relevant to URLLC.

For the nominal model, we take an in-depth look at wireless channel
characteristics such as spatial and temporal correlations based on
Jakes' model. Contrary to what has been claimed in the literature, we
find that standard Rayleigh fading processes are not bandlimited.
This has significant implications on the predictability of channels.
We also find that under reasonable conditions the spatial correlation
of channels provide a fading distribution that is not too far off from
an independent spatial fading model.
Additionally, we look at the impact of these channel models on
cooperative communication based systems. We find that while
spatial-diversity-based techniques are necessary to combat the adverse
effects of fading, time-diversity-based techniques are necessary to be
robust against unmodeled errors. Robust URLLC systems need to operate
with both an SNR margin and a time/repetition margin.

\end{abstract}
\vspace{-10pt}
\begin{IEEEkeywords}
\vspace{-10pt}
Fading processes, spatial-correlation, temporal-correlation, robustness, low-latency, high-reliability wireless, URLLC, industrial control, Internet-of-Things
\end{IEEEkeywords}

\bstctlcite{BIBcontrol}
\VerbatimFootnotes

\vspace{-10pt}
\section{Introduction}
\label{sec:intro}
Interactive and high performance applications demand strict communication requirements such as high-reliability and low-latency.
These can enable exciting applications like vehicle platooning, untethered AR/VR, exoskeletons, etc~\cite{6755599}.
Many of these applications are being envisioned to operate in the following fashion: tens to hundreds of nodes transmitting short messages of size tens to hundreds of bytes, that all need to reach their destinations in a few milliseconds with the probability that they violate their latency requirement (henceforth referred to as cycle time) being in the range of one-in-a-million ($10^{-6}$) to one-in-a-billion ($10^{-9}$)~\cite{Weiner}.

The biggest hurdle is that channel fading makes wireless channels fundamentally unreliable. To mitigate the effects of fading, diversity techniques exploiting time, frequency and space have been used successfully to build in reliability.
A message may be transmitted through multiple avenues -- be it retransmissions (time diversity), or coding across sub-carriers (frequency diversity), or using multiple antennas (spatial diversity).
As long as all avenues do not experience bad fades, the message will be successfully received.
However, the challenge lies in making reliable communication systems which satisfy low-latency requirements.
Essentially, \textbf{we need to find multiple avenues to transmit the message such that we can  guarantee there is at least one good path in a short time}.
This implies that time-diversity-based-techniques alone are not suitable for these low-latency, high-reliability applications as channels do not change fast enough to provide new fades reliably within the given latency bounds.
Frequency-diversity-based techniques suffer similar drawbacks as they would need channels to have high frequency-selectivity which in turn requires channels to reliably provide a large number of potentially good taps.
However, spatial-diversity-based techniques using multiple antenna communication do not seem to suffer the same drawbacks as time and frequency diversity to the same extent.
Essentially, as we will further establish in this paper, there is a high degree of confidence that wireless channels between nodes spread out in space are independently faded. Using these insights, cooperative communication based schemes have been developed to enable ultra-reliable low-latency communications (URLLC) and their performance has been studied~\cite{swamy2015cooperative, swamyTWC2017}.

Although spatial-diversity-based techniques seem to address the needs of URLLC, we need to take an in-depth look at wireless channel dynamics specifically in the low-latency regime.
It is imperative to understand the events that may occur in timescales ranging from tens of microseconds (corresponding to the length of a single short packet) to a few milliseconds (corresponding to the cycle time).
A knowledge of these events helps focus on issues that might otherwise be overlooked if a traditional quasi-static-channel model was considered. This in turn aids in designing systems that can sufficiently guard  against potentially adverse and rare events.
To illustrate the above point, let us consider the following scenario. There is a network of nodes that wants to send messages to their destinations wirelessly.
To guard against bad fading events, nodes nominate a buddy node to relay their message. Thus, each message gets two shots at getting to its destination -- first, directly from the source to the destination if the corresponding channel is good and second, through the buddy node (relay) if there were good channels between the source and the relay and the relay and the destination.
The source nominates the relay based on some past channel state information but the buddy node relays the information at \emph{a future time} -- say a millisecond later.
The traditional quasi-static-channel model suggests that channels remain static for a period of time given by the coherence time (which depends also on the system dynamics). Let us say that the traditional coherence time was $2$ms.
Given that the overall failure rate demanded is around $10^{-9}$, can the source be one-in-a-billion confident that $1$ms from now, that the relay will have a good channel to the destination? Do we trust the model that much?

Schemes that rely on the channel state information to maximize the data rate or choose `leader' nodes have been considered in the context of URLLC. In~\cite{jurdi2018variable}, the authors consider a centralized control system where the controller performs pilot-assisted channel estimation to adapt the transmission rate to each device based on the quality of its channel.
In~\cite{liu2018d2d}, the authors consider a leader selection scheme based on channel state information.
Both of these studies assume that the channels remain static for the duration of a cycle and have promising results. However, if channels could change much more rapidly than what the traditional coherence time would suggest, it could lead to an unmodeled degradation in performance for either scheme.




To resolve these questions, in this paper\footnote{Earlier versions of some of the results have appeared in~\cite{swamy2016robustness, swamy2018predicting}.}, we take a critical look at the main characteristics of wireless channels that impact the design and performance of different communication schemes. We do this by looking at both the nominal model of fading processes as well as identifying key dimensions of uncertainty to capture the impact of unmodeled effects. We then specifically analyze the impact of these channel characteristics on the performance of two cooperative communication schemes (`Occupy CoW' and `XOR-CoW') to illustrate the key pain points and reveal where margin needs to be added to the schemes to be able to absorb the effect of both nominal and unmodeled uncertainty.
Sec.~\ref{sec:related} reviews some of the related works in the field of wireless channel modeling.
The rest of the paper is organized as follows:
\begin{itemize}[
    \setlength{\IEEElabelindent}{\dimexpr-\labelwidth+8pt}
]
\item In Sec.~\ref{sec:channel_model} we study the temporal and spatial characteristics of wireless channels.
In Sec.~\ref{sec:temporal} we study wireless channel dynamics at two main time scales.\\
a) We study the channel dynamics on the order of tens to hundreds of microseconds which corresponds to the time duration of a short packet. We find that for short packets of duration under $100 \mu$s (motion under $0.01\lambda$), channels that are good enough stay quite static. There is no large variation in channel energy within this time if the channels started out to be good. \\ 
b) We study the channel dynamics on the order of hundreds of microseconds to a few milliseconds which corresponds to the cycle time as well as relaying events. We find that Rayleigh fading processes are not bandlimited. This has significant implications in channel quality prediction and relay selection techniques, but we only touch upon it in this paper.
\item In Sec.~\ref{sec:spatial} we study the spatial correlation of wireless channels and understand its impact on the fading distribution. We find that under reasonable conditions, we get fading distributions that are not too far off from an independent spatial fading model. In fact, a channel correlation bounded by $0.2$ (corresponding to nodes that are all at least $3\lambda$ apart as Eq.~\ref{eq:bessel} suggests) can easily be modeled as a drop in nominal SNR of just $0.05$dB.
\item Sec.~\ref{sec:impact} points out the key parameters used to bound unmodeled uncertainty --- the maximum probability of an unmodeled link error that is independent across transmitter/receiver pairs; the maximum probability of an unmodeled decoding error that is independent across time-slots, and the maximum probability of an unmodeled decoding error that compounds with the number of simultaneous transmitters, but is independent across time-slots\footnote{Space limits in the paper prevent our discussion of unmodeled errors that might be dependent across time, but independent across frequency channels. We just touch upon this in our concluding remarks since the analysis and philosophy is identical.}. Along with the improved nominal model from Sec.~\ref{sec:spatial}, this provides our robust model for wireless uncertainty.
\item Sec.~\ref{sec:protocol} summarizes the two main cooperative communication schemes being considered for the paper. In Sec.~\ref{sec:impact} we theoretically quantify the effect of channel dynamics on their performance. We find that these schemes can be made robust to most phenomena 
through the appropriate sizing of SNR margins as well as ``time footprint margins'' that result from the use of interleaved repetitions of messages.
Sec.~\ref{sec:conclusion} concludes the paper.
\end{itemize}


\vspace{-10pt}
\section{Related Work}
\label{sec:related}
Advancements in wireless communication technologies have been made possible by studies of channel characteristics in indoor and outdoor environments and extensive modeling of these environments~\cite{schaubach1992ray, anderson2002building}.
Characteristics of wireless channels like propagation loss in different environments have been instrumental in estimating coverage area for both traditional systems like cellular communication as well as newer technologies that leverage TV White Space bands~\cite{berg1992path, goldsmith2005wireless}.
Recently, wireless channel characteristics have been exploited for developing applications that leverage backscattering to image objects in the environment, building indoor positioning applications as well as novel low-power communication technologies~\cite{liu2013ambient}.
As advancements in hardware have enabled building mmWave radios, studying characteristics of wireless channels in indoor and outdoor environments in the mmWave band have become crucial to develop exciting applications including tether-less AR/VR and vehicle to vehicle (V2V) and vehicle to infrastructure (V2I) communication to enable autonomous vehicles~\cite{abari2016cutting, daniels200760, va2016millimeter, rappaport2015wideband}.
Most of these works have focused either on large scale statistics of channels (relevant for problems like capacity estimation) or dynamics in relatively larger scale for instance in the order of seconds~\cite{jakes1994microwave}.

By contrast, in this paper, we focus on temporal and spatial channel characteristics at both small as well as medium scales that capture various (potentially rare) events of interest that can ultimately cause a failure event. We also borrow from the practice within robust control of distinguishing between the nominal model and quantifying the uncertainty we have around that nominal model~\cite{zhou1998essentials}. This permits us to understand quantitatively how uncertainty about the wireless model itself translates into increased margins for key resources used (e.g.~SNR as well as the time-footprint of the protocol) by a wireless communication scheme so  that it can be robustly reliable to both the modeled and unmodeled uncertainty.

\section{Nominal Channel Models}
\label{sec:channel_model}
\begin{figure}[htb]
\begin{center}
\begin{subfigure}[b]{0.48\textwidth}
\centering
\includegraphics[width = \textwidth]{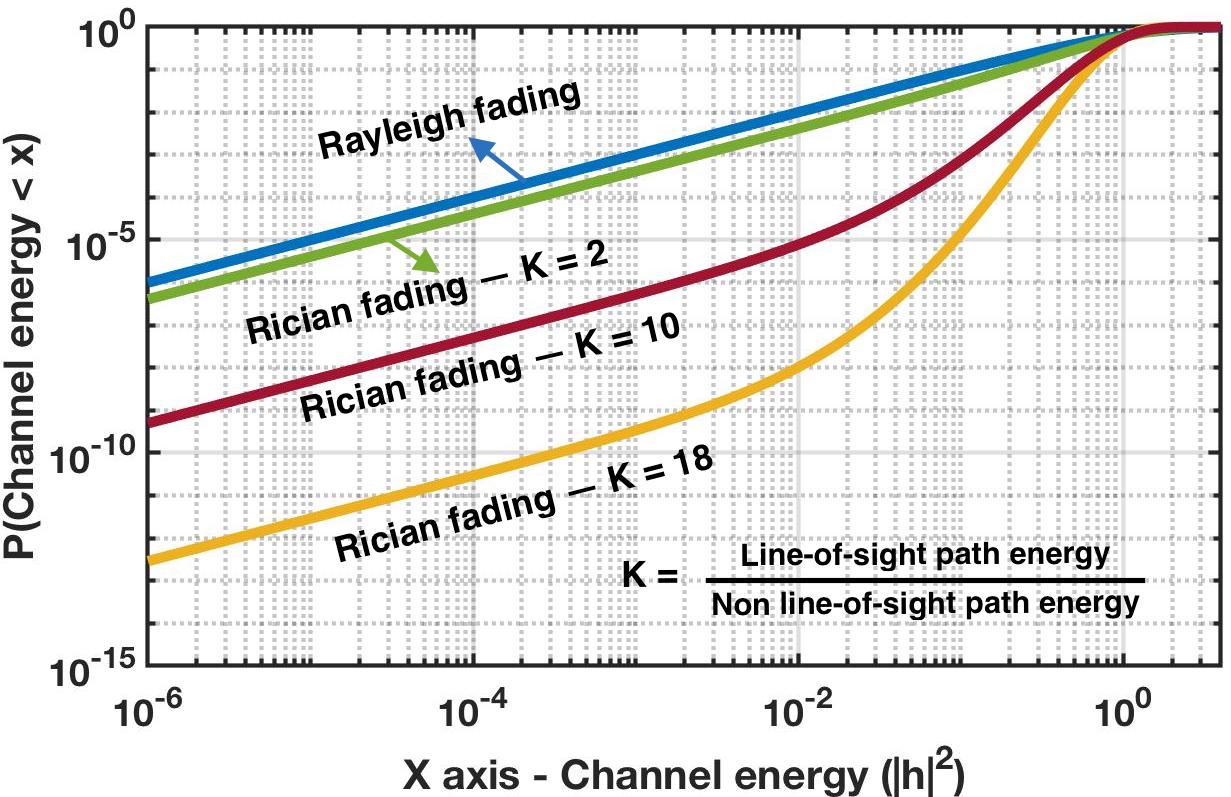}
\caption{The CDF of channel energy $|h|^2$ for Rayleigh (no line-of-sight path) and Rician (with line-of-sight path) faded channels for the same nominal SNR. The CDF for Rayleigh faded channels has more mass around $0$ (indicating deep fades) than Rician faded channels.}
\label{fig:channelCDF}
\end{subfigure}
~
\begin{subfigure}[b]{0.45\textwidth}
\centering
\includegraphics[width = \textwidth]{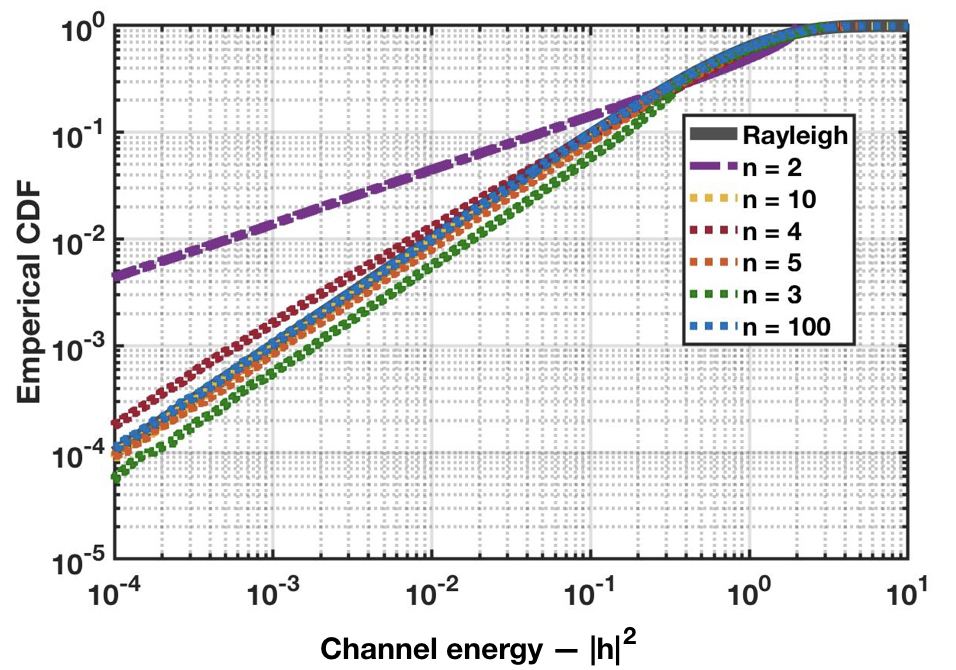}
\caption{The empirical CDF of channel energy $|h|^2$ for a Rayleigh faded channel obtained through the Jakes's model as given by Eq.~\eqref{eq:jake}. For different values of number of scatterers $n$ the empirical CDF lingers around the nominal CDF except at $n = 2$.}
\label{fig:empCDF}
\end{subfigure}
\caption{\small{Empirical and theoretical CDF of Rayleigh and Rician fades.}}
\end{center}
\vspace{-50pt}
\end{figure}
The main characteristics of wireless channels that are of interest in
the context of URLLC are the joint distributions of the channel fades
across time and space.
As mentioned earlier in Sec.~\ref{sec:intro} and \ref{sec:related},
most works in the literature have focused primarily on large-scale
statistics, on the order of tens of milliseconds to seconds.
The fading process is assumed to be bandlimited if the motions have bounded speed~\cite{jakes1994microwave}.
Moreover, often channels are assumed to be static for a duration of time given by the coherence time which is determined by how fast the nodes are moving etc.
However, it is imperative to question these assumptions and revisit these concepts keeping in mind that we are operating in low-latency regimes with short packets and aim to deliver high reliability.
Knowing which events can lead to failure (such as picking a relay assuming a static channel that actually ends up changing) and effectively addressing them by considering channel dynamics is key to building robust wireless communication systems for URLLC.


In this paper, we focus on studying Rayleigh fades to understand the worst-case scenario where there is no line-of-sight path in indoor environments. A line-of-sight path makes the fade distribution better (like Rician) as shown in Fig.~\ref{fig:channelCDF} where we see that
the mass around $0$ is higher in Rayleigh fading than in Rician fading
-- i.e., the chance of a Rayleigh faded channel being in a deep fade is higher than a Rician faded channel.

Rayleigh faded channels have traditionally been modeled using a sum-of-sinusoids like in Jakes's model~\cite{jakes1994microwave}.
We revisit the classical Jakes's model dynamics of Rayleigh faded channels and question whether the process is fundamentally bandlimited if the motions have bounded speed.
We consider only the effects of multipath as we focus on the {\em
  variations} at small timescales. The effects of shadowing,
diffraction, and other propagation effects can be divided into two
categories. One is nominal --- when we think about the nominal SNR in
this paper, this is the minimal SNR including all regions of shadowing and
normal path loss. Fundamentally, we are interested in modeling the situation
where every node can nominally hear every other node --- if there were
true ``dead spots'' where shadowing prevents this, this would
presumably be known ahead of time and would need to be dealt with using
``range-extension'' techniques. The other aspect of shadowing and
diffraction, namely transitions into and out of shadows, we fold into the unmodeled part of the channel that is addressed in Section~\ref{sec:impact}.

\begin{figure}
\centering
\includegraphics[width = 0.3 \textwidth]{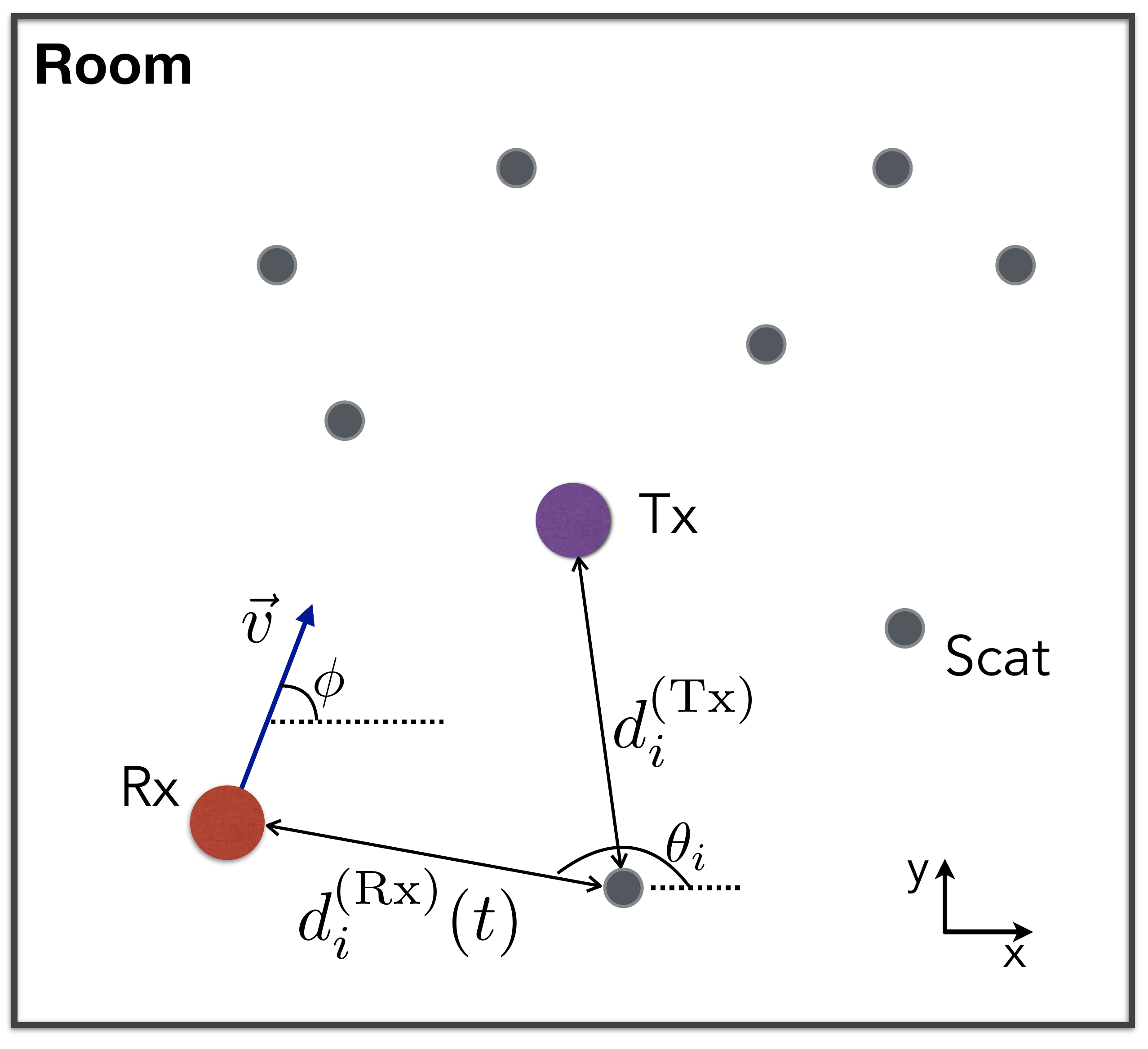}
\vspace{-10pt}
\caption{\small{Room setup with $n$ static scatterers, a static transmitter and a mobile receiver.}}
\label{fig:room}
\vspace{-40pt}
\end{figure}
Consider a two-dimensional room with $n$ static scatterers\footnote{In
  Fig.~\ref{fig:empCDF}, we also see that the convergence is very fast
  with the number $n$ of scatterers. Essentially, it has converged by
  $n=3$. However, $n=2$ needs to be considered as a special case if
  the deployment environment is one in which that could happen. We do
  not dwell on that case here, but in effect, its CDF tells us that we
  need to pay an extra $10+$dB of transmit power if we want to avoid
  deep fades in the context of two-scatterer environments. This is
  because the case of exactly two scatters results in nulls along long
  valleys where the two paths have canceling phases. Once we have
  three or more paths, valleys of nulls can no longer exist ---
  instead, we get isolated nulls. This geometric fact tells us why the
  convergence is so fast.} distributed uniformly at random.
Let there be a static single-antenna transmitter in the middle of the room and a single-antenna mobile receiver moving at a constant speed $v$ in some random direction inside the room (illustrated in Fig.~\ref{fig:room}).
Let the transmitter be transmitting a tone at frequency $f_c$ (wavelength $\lambda_c$).
The channel coefficient between the transmitter and the receiver at any time $t$ is given by
\begin{equation}
    h(t) = \frac{1}{\sqrt{n}} \sum_{i = 1}^{n} \exp{\left(j \frac{2\pi (d^{(\textrm{Rx})}_i(t) + d^{(\textrm{Tx})}_i)}{\lambda_c}\right)}\\
\label{eq:jake}
\end{equation}
where $d^{(\textrm{Rx})}_i(t)$ is the distance of the scatterer $i$ from the receiver at time $t$ and $d^{(\textrm{Tx})}_i$ is the distance of the scatter $i$ from the transmitter (both are assumed to not be moving for simplicity). The $1/\sqrt{n}$ normalization in Eq.~\eqref{eq:jake} is to keep the marginal variance the same across different numbers $n$ of scatterers. This is because our goal is to understand the reliability impacts of the variability that fading brings --- this tells us how much higher we need to make the nominal SNR to be able to absorb the impact of these fades without losing system-level reliability.
Eq.~\eqref{eq:jake} follows from the results in~\cite{pop2001limitations}.

\vspace{-10pt}
\subsection{Channel variation as a Gaussian process}
We want to understand how channels between a pair of antennas vary as one (or both) antennas move while the environment (scatterers) remains largely stationary.
This model captures the small-scale variations that we are interested in, where the nodes are moving at a reasonable speed but for small amounts of time (in ms).
The channel coefficient at any point in time is marginally distributed as a complex normal (as the CLT suggests for the expression in Eq.~\eqref{eq:jake}), and the channel coefficient process through time can be modeled as a Gaussian process.
The parameters that we need to define the Gaussian process are the means and the covariance functions which depend on the distance that the receiver has moved.
We assume that the velocity $\vec{v}$ of the receiver is constant over the time durations of interest such that the position of the receiver $\vec{s}(t)$ at time $t$ is given by
\begin{equation}
\vec{s}(t) = \vec{s_0} + \vec{v}t = (x_0 + vt \cos{\phi}, y_0 + vt \sin{\phi})
\end{equation}
where $\vec{s_0} = (x_0, y_0)$ is the initial position of the receiver at $t = 0$ (uniformly distributed in the room), $\phi$ is the angle of motion of the receiver with respect to the $x$-axis (uniformly distributed over $[0, 2\pi)$).
Let the position of scatterer $i$ be given by $\vec{s_i} = (x_i, y_i)$.
The distance of the receiver from scatterer $i$ at time $t$ is given by
\begin{align}
    &d^{(\textrm{Rx})}_i(t)   = \|\vec{s}(t) -  \vec{s_i}\| = \sqrt{(x_0 + vt \cos{\phi} - x_i)^2 + (y_0 + vt \sin{\phi} - y_i)^2}\nonumber \\
    &= \sqrt{d^{(\textrm{Rx})}_i(0)^2 + (vt)^2 + 2vtd^{(\textrm{Rx})}_i(0)\cos{(\theta_i - \phi)}}
    \label{eq:distance}
\end{align}
where $d^{(\textrm{Rx})}_i(0)$ is the distance of the receiver from the scatterer $i$ at $t = 0$ and $\theta_i$ is the angle made by the line joining the scatterer and the receiver at time $t = 0$ which is independent of $\phi$.

\begin{figure}
\begin{center}
\begin{subfigure}[b]{0.45\textwidth}
\centering
    \includegraphics[width = \textwidth]{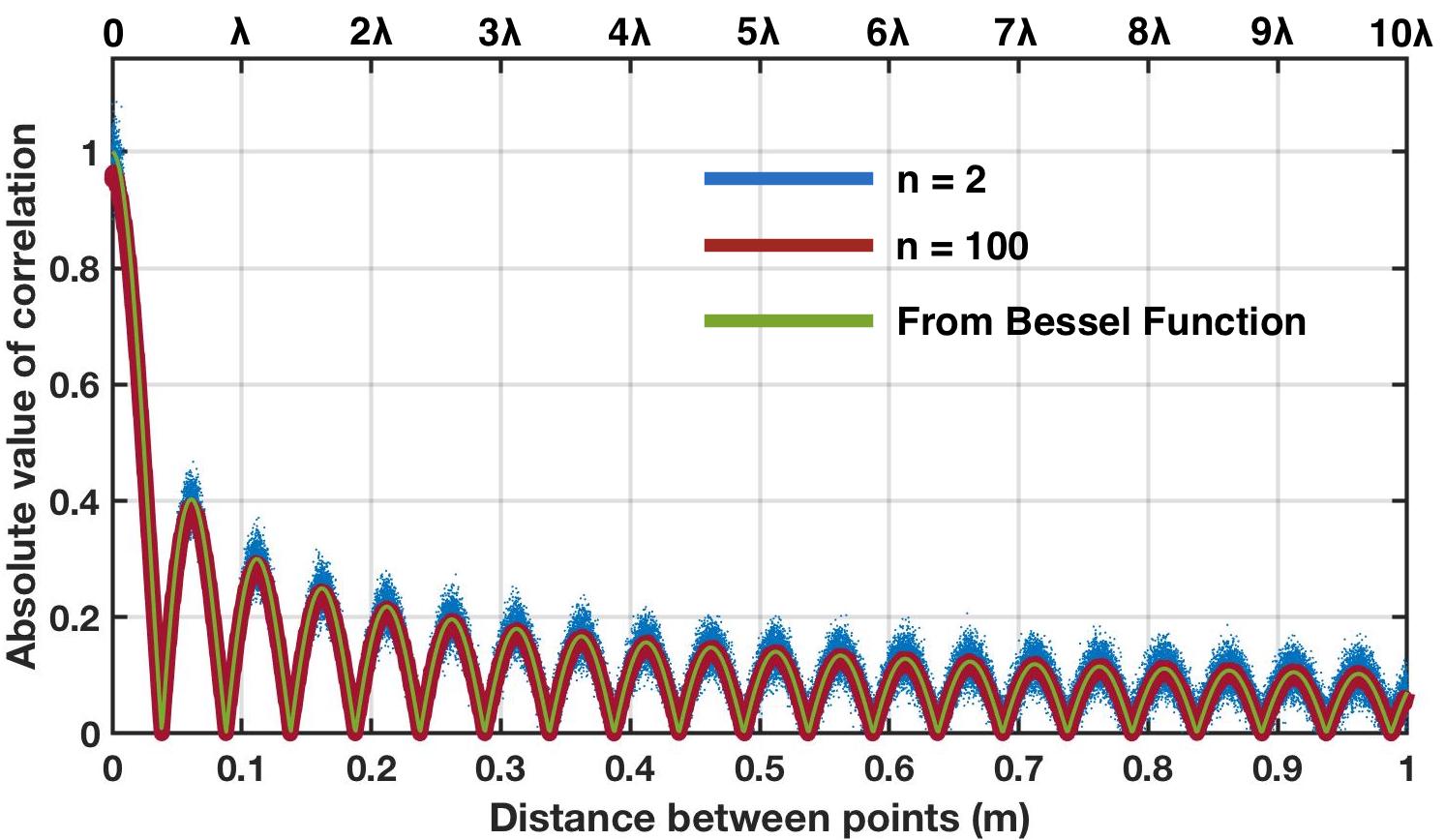}
\caption{{Simulated absolute value of the cross-covariance of the fading process for $f_c = 3$GHz. The curves are exactly as predicted by Eq.~\ref{eq:bessel} as the red curve corresponding to $n = 100$ aligns exactly with the green curve which corresponds to the actual Bessel function.}}
\label{fig:bessel}
\end{subfigure}
~
\begin{subfigure}[b]{0.45\textwidth}
\centering
\includegraphics[width = \textwidth]{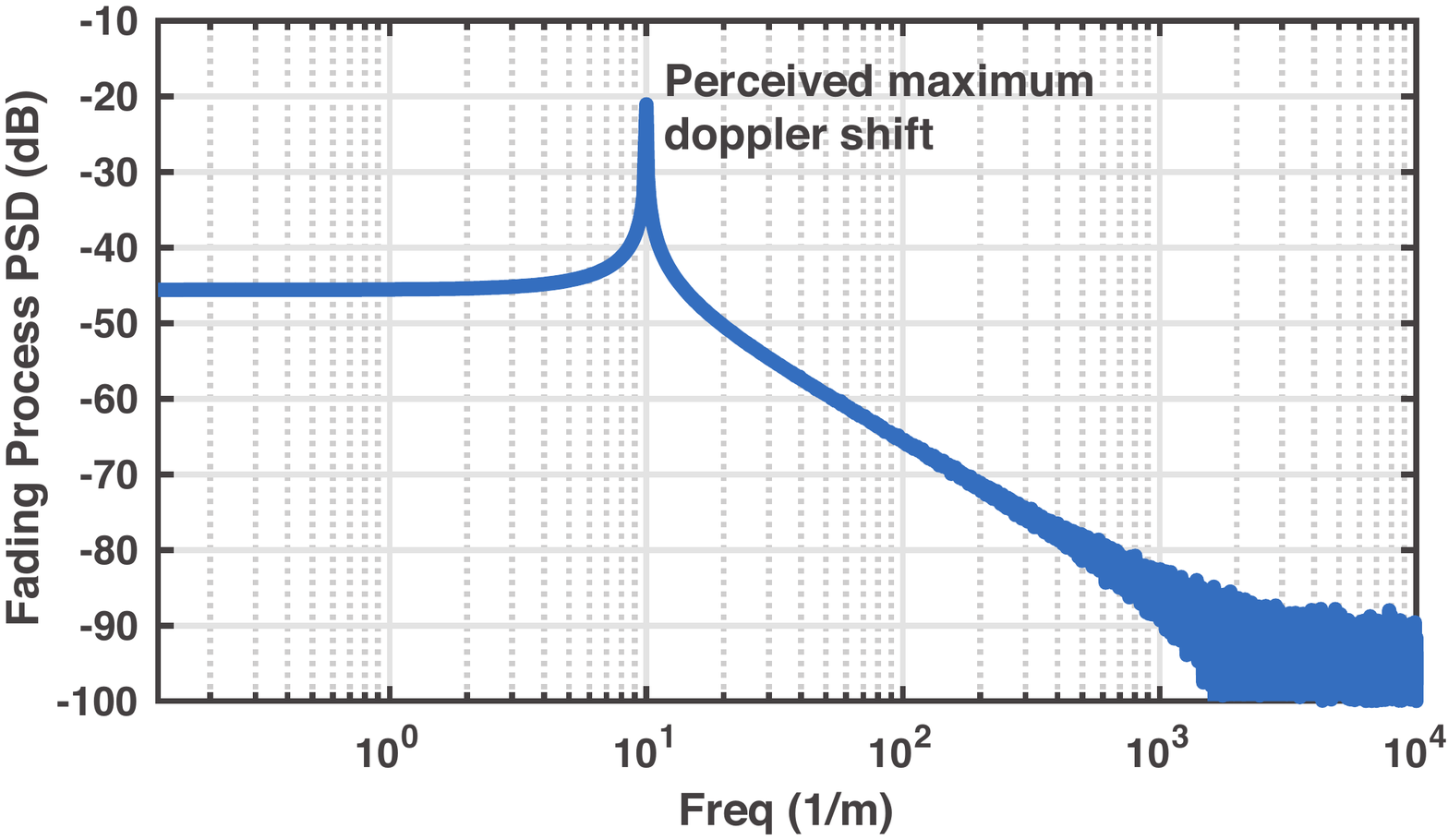}
\caption{{One-sided power-spectral density (PSD) of the fading process for $f_c = 3$GHz. Traditionally the PSD has considered to exist only until the perceived maximum Doppler shift but there clearly exists energy beyond this frequency and therefore the process can be considered to not be bandlimited.}}
\label{fig:psd}
\end{subfigure}
\caption{\small{Fading process and its cross-covariance and PSD.}}
\end{center}
\vspace{-50pt}
\end{figure}

As we are interested in channel dynamics and correlations, it is natural to examine the covariance of the in-phase, $\Re{(h(t))}$ and the quadrature components, $\Im{(h(t))}$ of the wireless channel as a function of speed $v$ and time $t$. Let us denote this covariance by $k(v,t) = \mathbb{E}[\Re{(h(t))}\Re{(h(0))}] = \mathbb{E}[\Im{(h(t))}\Im{(h(0))}]$. Let us also look at the cross-covariance of the channel coefficient ($h(t)$) given by $\tilde{k}(v,t) = \mathbb{E}[h(t)h^*(0)]$. Since the in-phase and quadrature components are uncorrelated (verified through simulations), $\tilde{k}(v,t) = \mathbb{E}[\Re{(h(t))}\Re{(h(0))}] + \mathbb{E}[\Im{(h(t))}\Im{(h(0))}] = 2k(v,t)$. We now calculate ${k}(v,t)$ through $\tilde{k}(v,t)$.
We have,

{\small
\begin{align}
&\tilde{k}(v,t) = \mathbb{E}[h(t)h^*(0)] \nonumber \\
&=\frac{1}{n}\mathbb{E}\Bigg[ \sum_{i = 1}^{n} \exp{\left( j \frac{2\pi}{\lambda_c} \left(d^{(\textrm{Rx})}_i(t) - d^{(\textrm{Rx})}_i(0)\right)\right)} + \sum_{i \ne j} \exp{\left( j \frac{2\pi}{\lambda_c} \left(d^{(\textrm{Rx})}_i(t) - d^{(\textrm{Rx})}_j(0) + d^{(\textrm{Tx})}_i - d^{(\textrm{Tx})}_j\right)\right)} \Bigg].
\label{eq:cov}
\end{align}
}%
As $d^{(\textrm{Rx})}_i(t)$ and $d^{(\textrm{Rx})}_j(0)$ are independent for $i \ne j$ and the scatterers are distributed uniformly across the room, the expectation of the cross term in Eq.~\eqref{eq:cov} is $0$.
\begin{eqnarray}
\begin{aligned}
\tilde{k}(v,t)
&= \mathbb{E}\left[ \exp{\left( j \frac{2\pi}{\lambda_c} \left(d^{(\textrm{Rx})}_i(t) - d^{(\textrm{Rx})}_i(0)\right)\right)} \right]
\label{eq:cov-final}
\end{aligned}
\end{eqnarray}
Substituting Eq.~\eqref{eq:distance} in \eqref{eq:cov-final}, for small movements ($\frac{vt}{d_i} \approx 0$), the covariance function is given by
\begin{equation}
\tilde{k}(v,t) = J_0\left( \frac{2 \pi }{\lambda_c} vt \right)
\label{eq:bessel}
\end{equation}
where $J_0(\cdot)$ is the Bessel function of the first kind (also derived in~\cite{gans1972power}).
Fig.~\ref{fig:bessel} shows the simulated absolute value of the expected cross-covariance of the fading process as a function of distance which matches Eq.~\eqref{eq:bessel}. The convergence is rapid with the number of scatterers $n$.

\vspace{-10pt}
\subsection{Bandwidth of fading processes}
The power spectral density of the complex fading process has indeed been looked at in studies like~\cite{gans1972power, jakes1994microwave, baddour2005autoregressive}.
However, they make an essential assumption: that the power spectrum is bowl shaped and the contribution of frequencies higher than the perceived maximum frequency is zero -- essentially, the fading process is bandlimited.
The unilateral Laplace transform of the Bessel function ($\mathcal{L}(J_0(x)) = 1/\sqrt{1 + s^2})$ has poles on the imaginary axis.
Therefore, the Fourier transform gets tricky -- how do we deal with these poles? Studies so far (such as~\cite{gans1972power, jakes1994microwave, baddour2005autoregressive}) seem to have elected to restrict the Fourier transform of the Bessel function until the `maximum' Doppler shift (i.e., $v/\lambda_c)$, possibly to address these poles.
However, our simulations show that the fading process is not bandlimited -- it has energy even beyond the traditionally assumed maximum Doppler shift. This surprising discovery was also supported by looking numerically directly at the Bessel function.
The standard assumption of ignoring the response outside the maximum Doppler shift was reasonable when the focus was on the average or typical behavior of the process.
However, for URLLC we are interested in rare events with probabilities on the order of $10^{-9}$ so taking this into account is important.

\begin{figure}
\begin{center}
\begin{subfigure}[b]{0.45\textwidth}
\centering
\includegraphics[width = \textwidth]{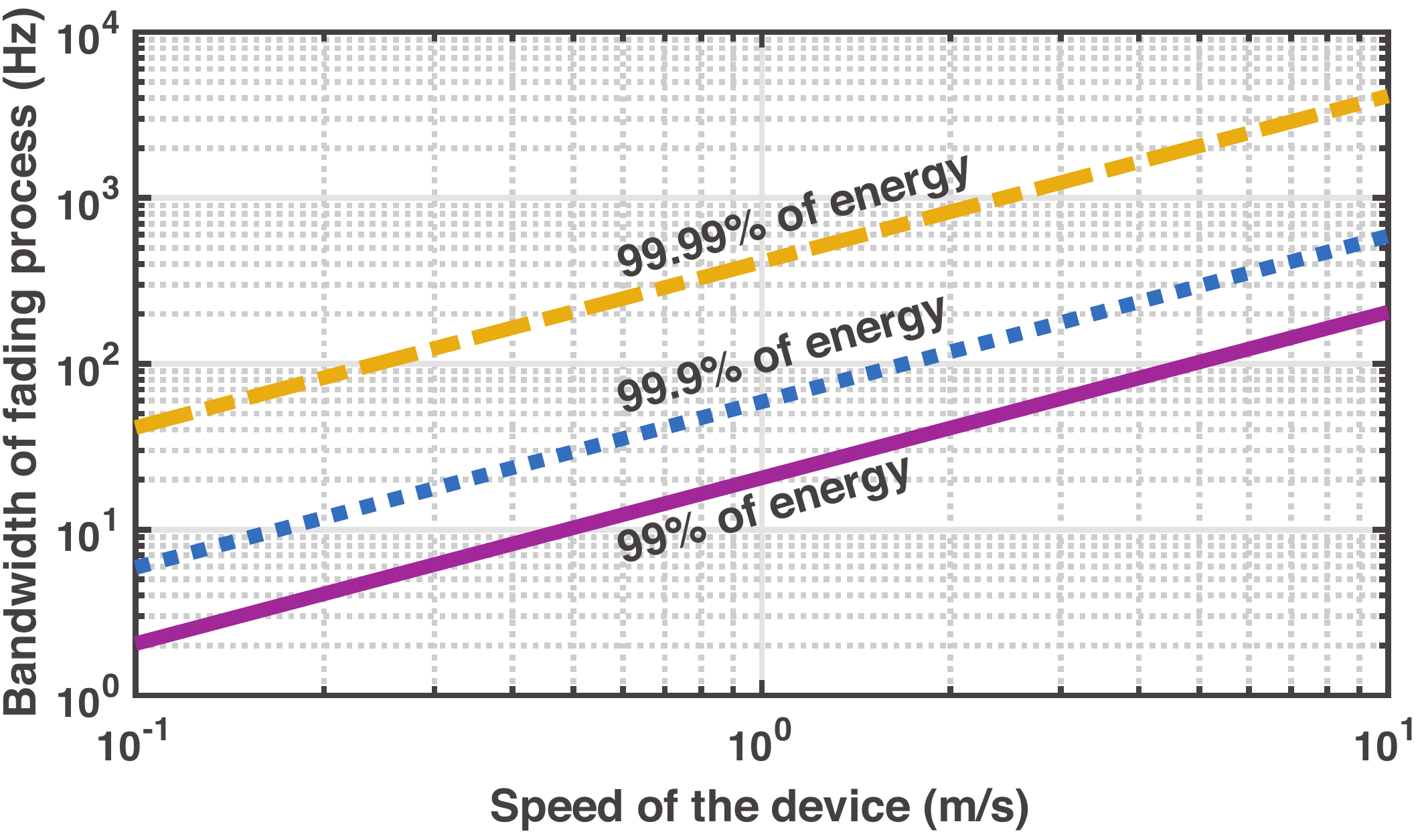}
\caption{{Bandwidth around the center frequency which captures $99 \%$, $99.9 \%$ and $99.99 \%$ of the energy of the fading process as a function of node speed}.}
\label{fig:bandwidth}
\end{subfigure}
\hspace{10pt}
\begin{subfigure}[b]{0.48\textwidth}
\centering
\includegraphics[width = \textwidth]{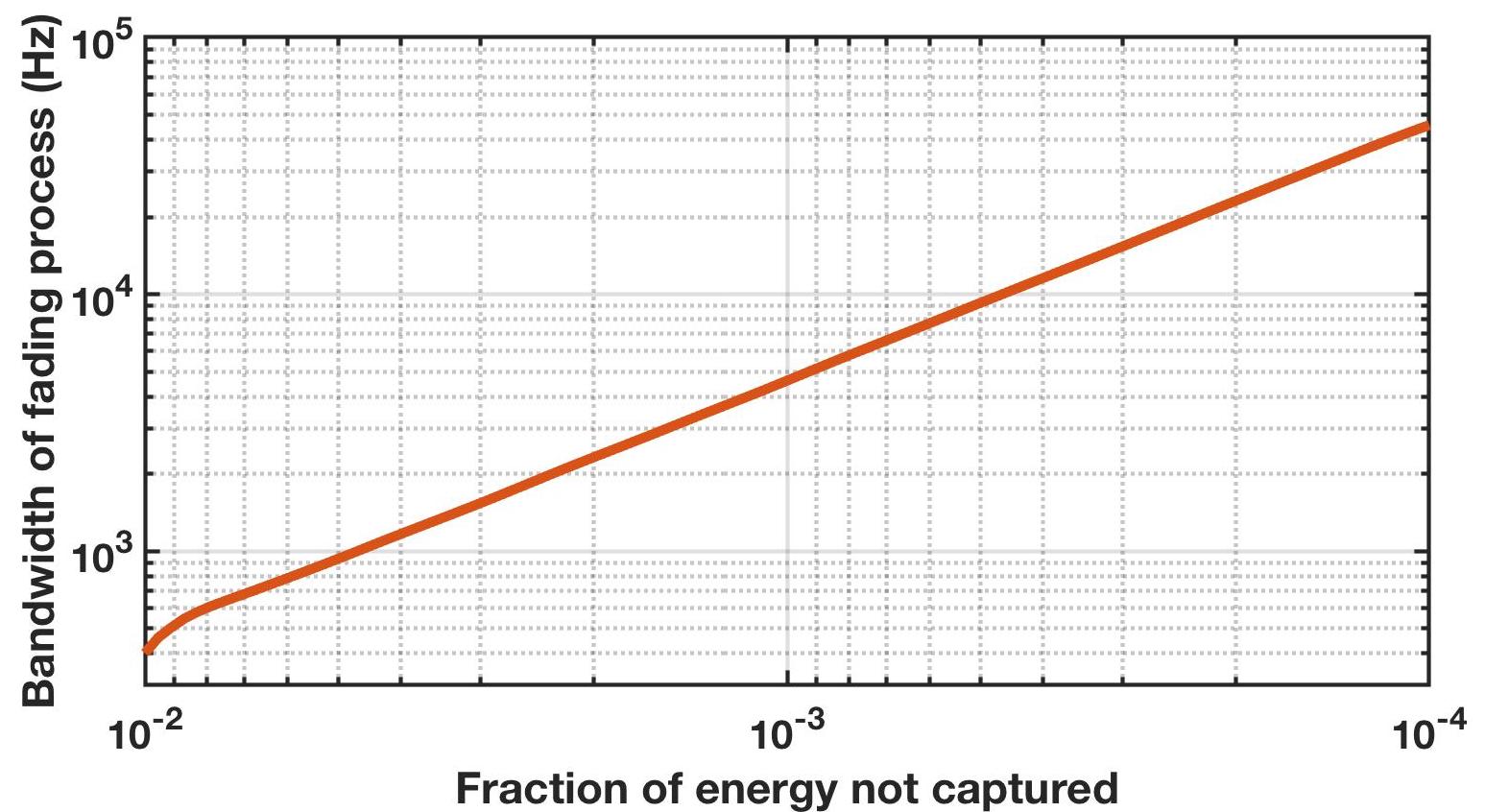}
\caption{{Bandwidth around the center frequency which captures varying amount of energy of the fading process while the node speed is kept constant at $10$m/s}.}
\label{fig:bandwidth_v_error}
\end{subfigure}
\end{center}
\caption{\small{Relationship between the bandwidth of the fading process and the energy it contains.}}
\vspace{-40pt}
\end{figure}
Figure.~\ref{fig:psd} plots the one-sided power spectral density (obtained through simulations) of the fading process, for center frequency $f_c = 3$GHz.
We do see the bowl shape that is traditionally expected until the spatial frequency of $10$/m (corresponding to the maximum Doppler frequency) but it clearly doesn't die down to $0$ immediately beyond the maximum Doppler frequency, instead decaying at the rate of $20$dB per decade.

Traditionally, the bandwidth of a process has been characterized through the amount of energy in it. Fig.~\ref{fig:bandwidth} plots the bandwidth that contains $99 \%$, $99.9 \%$ and $99.99 \%$ of the energy for various node speeds for center frequency $f_c = 3$GHz.
We see the expected linear scaling with speed but we also see the increase in bandwidth with increasing energy content. Specifically, Fig.~\ref{fig:bandwidth_v_error} plots the increase in bandwidth with increasing energy captured in it where we see that the $10$dB/decade increase is consistent with the behavior in Fig.~\ref{fig:psd} (where there is a $20$dB/decade drop and the factor of two comes because of the squaring effect).

\vspace{-10pt}
\subsection{Temporal characteristics of wireless channels}
\label{sec:temporal}
In the URLLC context, there are two time scales of interest -- the duration of a single short packet and the overall cycle time. Therefore, we focus on the temporal characteristics at these two time scales to capture the variations within a packet duration and variations within a cycle.

\subsubsection{Channel variations within a packet duration}
\label{sec:temporalpacket}
As we are interested in small messages (payloads), the corresponding packet durations will also be small. For payloads of sizes $10$s to $100$s of bytes, the packet duration is at most $50\mu$s long if we assume the data-rate is on the order of $20$Mb/s. Let us consider that a node may move at a maximum speed of $10$m/s. How does the channel energy ($|h|^2$) change over the duration of a packet in the above setup?

\begin{figure}[h]
\centering
\includegraphics[width = 0.48\textwidth]{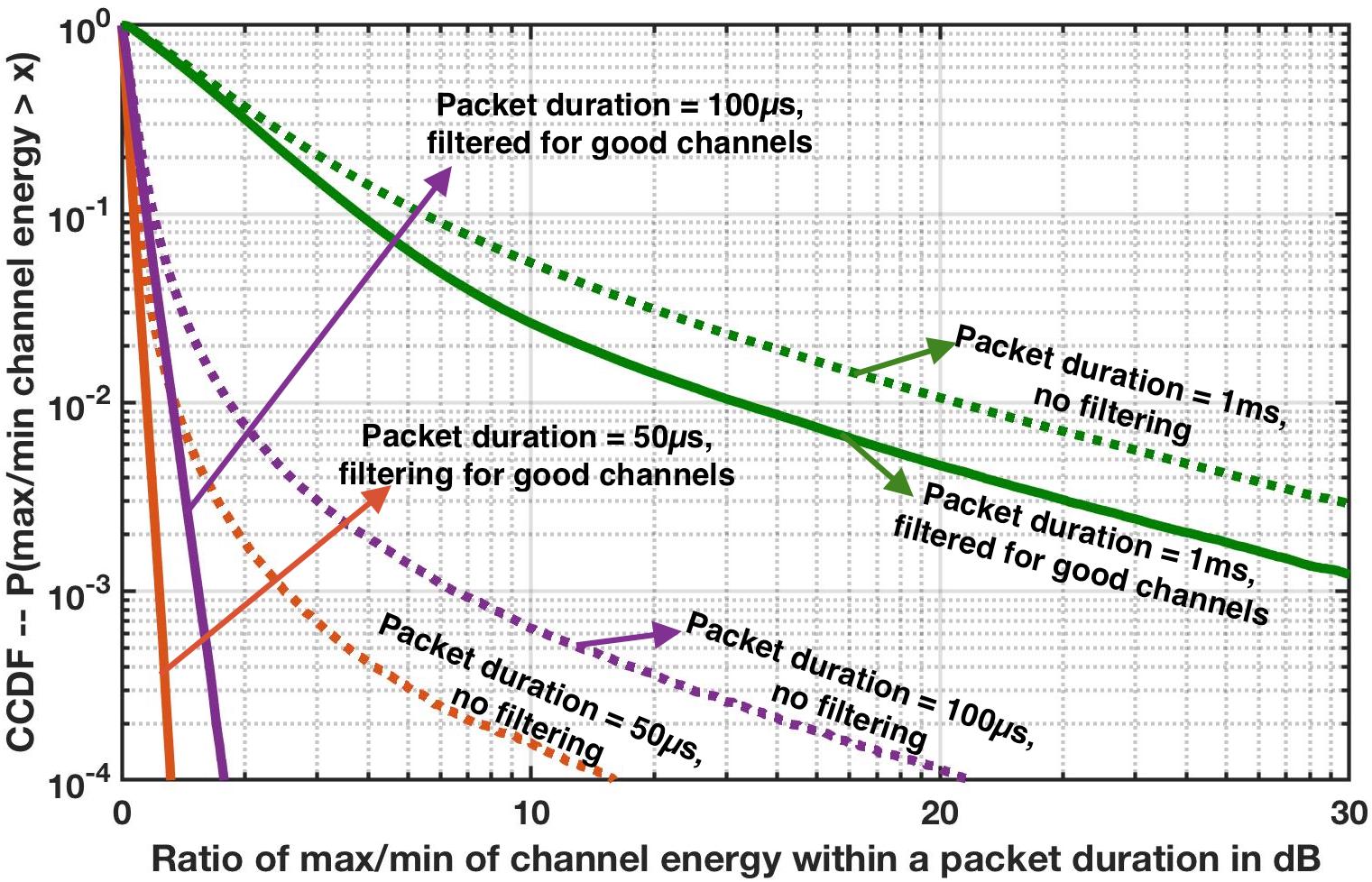}
\vspace{-10pt}
\caption{\small{CCDF of the ratio of max channel energy by min channel
    energy in dB within a packet duration for various packet
    durations. The receiver is traveling at a speed of $10$m/s and the
    center frequency is $3$GHz. The dotted curves correspond to all
    channels and the solid curves correspond to those channels that
    are good at the beginning of the packet. For short packet sizes of
    $50 \mu$s, there is no discernible change in the channel energy
    when conditioned on the initial channel being good but as the packet duration increases, we see  bigger variations become possible.}}
\label{fig:quasi-static}
\vspace{-40pt}
\end{figure}

Fig.~\ref{fig:quasi-static} studies the variation of channel energy within a packet for a static transmitter transmitting at center frequency of $3$Ghz and a mobile receiver moving at speed $10$m/s.
The orange dotted line corresponding to packet duration of $50\mu$s plots the CCDF of the ratio of maximum channel energy by the minimum channel energy within the given packet duration and the variation seems high.
Does this mean that channel energy varies so wildly -- the channel energy can fall by more than $10$dB within a packet? The answer is: it depends. If we condition to only look at channels that are good channels (energy above $-7$dB) at the beginning of the packet, then the variation is extremely minimal (less than $1$dB) -- as shown by the solid orange curve. This means that a good channel will remain reliably good for short packet sizes. The huge variations in channel energy were due to already badly faded channels -- even small variations manifest in a big way.

However, the story is very different for medium to long packet sizes. We look at packets $100\mu$s and $1$ms long and see that even after conditioning on looking at the channels that started out good, there is a \emph{significant} variation in their energy $1$ms later (the effect is less pronounced for packet duration $100\mu$s). The traditional coherence time for this setup is $2.5$ms. However, Fig.~\ref{fig:quasi-static} shows that even good channels do not reliably remain static for $1$ms. This suggests that in the context of URLLC, having small packets (on the order of $10\mu$s) can guarantee better stability.
This also has significant implications for channel prediction and relay selection as studied in~\cite{swamy2018predicting}.


\subsubsection{Channel variations within a cycle}
\label{sec:temporalcycle}
We have looked at variation in channels for both small and large packet sizes and saw that for large packet sizes $\approx 1$ms, the channel varies significantly. Therefore, simply assuming that channels remain constant for the traditional coherence time duration could potentially lead to severe degradation in performance.
As described in Sec.~\ref{sec:intro}, there are various schemes and techniques such as relay and leader selection and transmission rate optimization for which knowing the channel variation on the scale of milliseconds is critical. In this section, we study the variations of channels on the order of milliseconds and specifically look at the predictability of channels as a key measure in the context of URLLC.

\begin{figure}
\begin{center}
\begin{subfigure}[b]{0.48\textwidth}
\centering
\includegraphics[width = \textwidth]{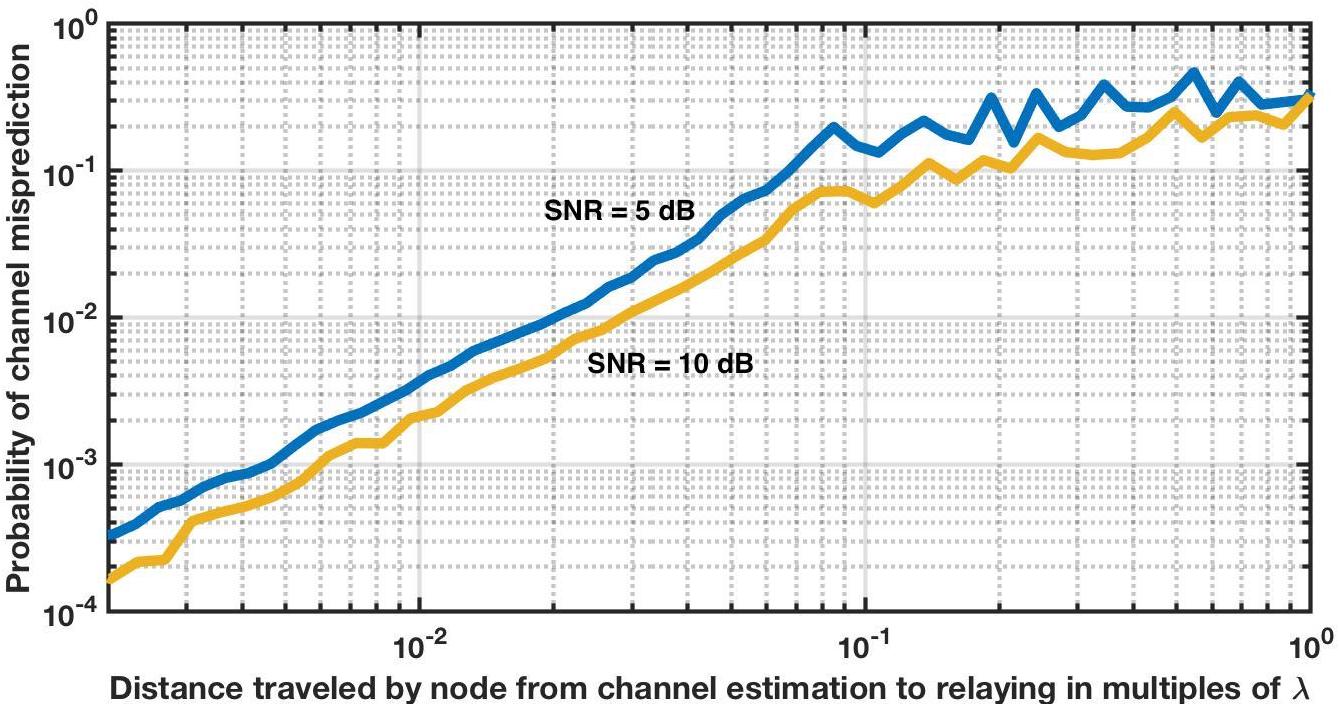}
    \caption{{The probability of channel quality misprediction as a
        function of how far the node has moved (at 10m/s) since the
        last channel measurement. This figure was obtained for
        predicting the channel using the past $3$ms of channel
        measurements taken every millisecond, and nominal SNRs $5$ and $10$dB.}}
\label{fig:coherenceTime}
\end{subfigure}
~
\begin{subfigure}[b]{0.48\textwidth}
\centering
\includegraphics[width = \textwidth]{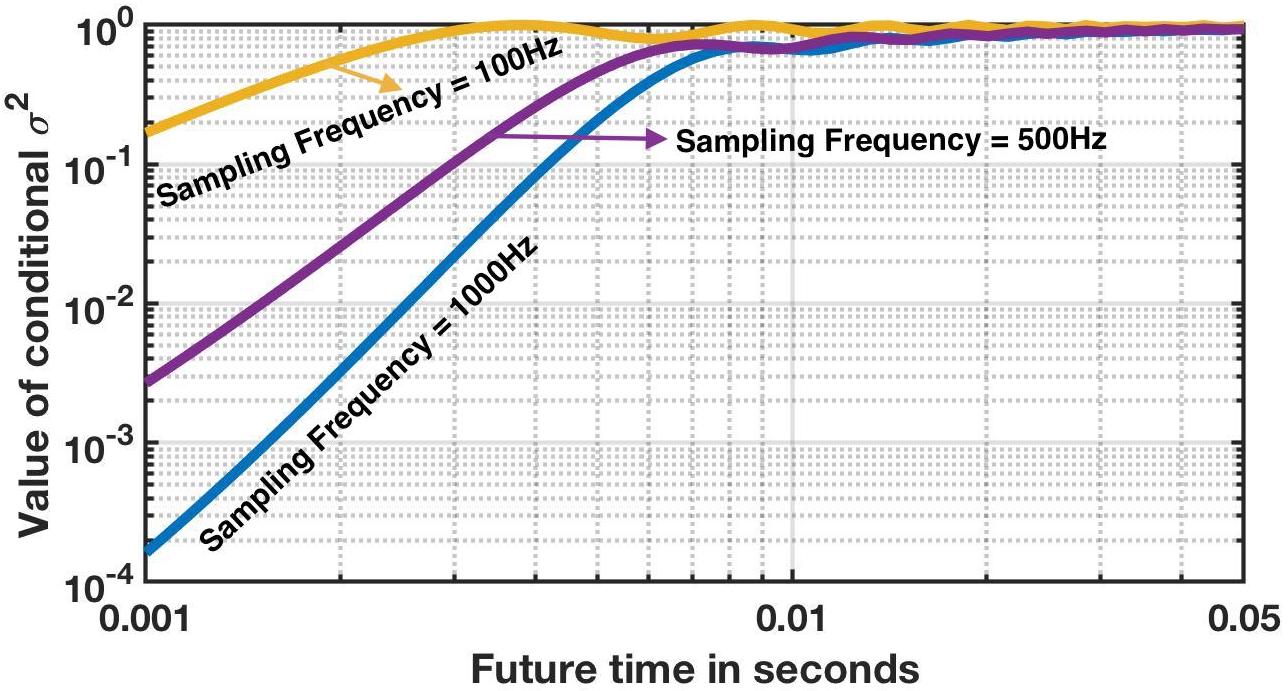}
    \caption{{The conditional variance of the channel energy
        distribution predicted by Eq.~\eqref{eq:variance} as a
        function of future time and sampling frequency of the channel
        coefficient. The higher the sampling rate, the lower the
        variance. The farther out in the future, the closer the
        variance becomes to the unconditional variance.}}
\label{fig:sigma_sq}
\end{subfigure}
\caption{\small{The effects of channel changing within a cycle through the perspective of future channel characteristics.}}
\end{center}
\vspace{-50pt}
\end{figure}

Consider the channel between a pair of nodes, say a source and a
destination. Via feedback and/or reciprocity, suppose the source has knowledge of past channels given by $\vec{h} = \begin{bmatrix} h_1 & h_2 & \hdots & h_m \end{bmatrix}^T$from times $\vec{t} = \begin{bmatrix} t_1 & t_2 & \hdots & t_m \end{bmatrix}^T$. We want to find the distribution of $h_{m+1}$ at time $t_{m+1}$ conditioned on $\vec{t}$ and $\vec{h}$. We assume that the channel coefficient variation is a Gaussian process, and use simple linear estimation. Therefore, $\lbrace\vec{h}, h_{m+1} \rbrace$ form a multivariate normal and the distribution of $h_{m+1}$ conditioned on $\vec{h}$ is a complex normal distribution.
Let $$\mathbf{K} = \begin{bmatrix} k(v, t_1 - t_1) & k(v, t_2 - t_1) & \hdots & k(v, t_m - t_1)\\
k(v, t_2 - t_1) & k(v, t_2 - t_2) & \hdots & k(v,t_m - t_2) \\
\vdots & \vdots & \ddots & \vdots \\
k(v,t_m - t_1) & k(v,t_m - t_2) & \hdots & k(v,t_m - t_m)
\end{bmatrix}$$
be the covariance matrix of the in-phase and quadrature components corresponding to the times of observations so far. Let
$\mathmbox{\mathbf{K_*} = \begin{bmatrix} k(v,t_{m +1} - t_1) & \hdots & k(v,t_{m+1} - t_m) \end{bmatrix}}$
be the covariance matrix of the in-phase and quadrature components corresponding to the future time of interest and the times of observations so far.
Also, let $\mathmbox{\mathbf{K_{**}} = \left[ k(v, t_{m+1} - t_{m+1})\right] = \left[ k(v, 0)\right]}$ be the variance of the in-phase and quadrature components.
Let $\vec{h}_I = \text{Re}\lbrace\vec{h}\rbrace$ be the vector of the in-phase components of $\vec{h}$ and $\vec{h}_Q = \text{Im}\lbrace\vec{h}\rbrace$ be the vector of the quadrature components of $\vec{h}$.
Then, the mean of the distribution of the in-phase $\mu_I$ and the quadrature component $\mu_Q$ of $h_{m+1}$ conditioned on $\vec{t}$ and $\vec{h}$ is given by
\begin{align}
\mu_I = \mathbf{K_*} \mathbf{K^{-1}} \vec{h}_I,\hspace{20pt}
\mu_Q = \mathbf{K_*} \mathbf{K^{-1}} \vec{h}_Q.
\label{eq:mean}
\end{align}
The conditional variance of both the in-phase and quadrature components is given by
\begin{equation}
\sigma_{c}^2 = \mathbf{K_{**}} - \mathbf{K_*} \mathbf{K^{-1}}\mathbf{K_*}^T.
\label{eq:variance}
\end{equation}
As mentioned earlier, the goodness or the quality of a channel is captured by the energy ($|h|^2$) in the channel.
The conditional distribution of the energy of the channel $|h_{m+1}|^2$ is given by
\begin{equation}
|h_{m+1}|^2 \sim \text{Rice}(\nu, \sigma_c)
    \label{eq:rice}
\end{equation}
where $\nu = \sqrt{\mu_I^2 + \mu_Q^2}$, $\sigma$ is given by
Eq.~\eqref{eq:variance} and $\text{Rice}(\nu,\sigma_c)$ is the Rician
distribution with parameters $\nu$ and $\sigma_c$. The value of
$\sigma^2_c$ is a direct indicator of the variability of the channel
at the future time $t_{m+1}$. In addition to the distance into the future, the value of $\sigma^2_c$ crucially depends on how fast we are sampling the channel as seen in Fig.~\ref{fig:sigma_sq}.

\noindent \textbf{Redefining coherence time/distance:}\\
We are interested in using the above model to understand the predictability of wireless channels.
Traditionally, channels have been considered to be static for a period of time or distance dictated by the system dynamics and carrier frequency. This is the notion of coherence time or distance. This has been a good rule-of-thumb for traditional cellular or WiFi-type systems as they focus mainly on average performance.
However, URLLC requires guarantees on worst-case performance which challenges the traditional notion of coherence time or distance. As seen earlier in Sec.~\ref{sec:temporalcycle}, channels simply do not stay static for the traditional coherence time.
Essentially, given some past measurements of a channel, we can predict the channel energy in the future time or distance. If the channel energy is greater than some threshold (dictated by the nominal SNR), we label that to be a good channel. However, there is a probability of mislabelling the channel (similar in spirit to demodulation error) which crucially depends on the \emph{future time or distance} through $\sigma^2_c$.
This error in mislabelling the channel is the fidelity corresponding to the future time or distance.
The nearer the future time is, the lower the misclassification probability would be and the farther out the future time is, the higher the misclassification probability would be.
To this end, {\bf we propose a more nuanced notion of coherence time or distance: the time or distance over which a channel is predictable to a given reliability.}

Fig.~\ref{fig:coherenceTime} shows the distribution of coherence
distance in units of wavelengths for a single channel. This was
obtained by considering predicting a channel distributed as
Eq.~\eqref{eq:rice} to be good or bad when operating at nominal SNRs
of $5$ and $10$dB. We see that the prediction is incorrect about $0.2
\%$ of the time even when the node has moved only
${\frac{1}{100}}^{\text{th}}$ of the wavelength. The rule-of-thumb is
that for every order of magnitude in distance, the probability of
error goes up by about $1.5$ orders. It plateaus around the
unconditional outage probability. If a node travels sufficiently far,
say $\frac{\lambda}{4}$ -- it will have little channel correlation from where it began.

\vspace{-10pt}
\subsection{Spatial characteristics of channels}
\label{sec:spatial}
We have focused primarily on the temporal characteristics of wireless channels. As spatial-diversity-based protocols are promising candidates for enabling URLLC, it is essential to understand the spatial correlation of wireless channels. If we end up in a scenario where channels are heavily correlated, then spatial-diversity-based schemes may fail.
Our investigations in Sec.~\ref{sec:channel_model} as well as experimental evidence show that \textbf{wireless channels are spatially correlated}. What does this mean for the schemes that want to exploit spatial diversity? Is this a recipe for disaster or is the degradation actually something manageable?
The answer to this question depends on how far apart the nodes are.
Essentially, there are two scenarios: one (unrealistic and impractical) where all nodes are within a wavelength apart from each other and the other (realistic and practical) where nodes are reasonably spread out in the environment.\\
\textbf{\emph{Case 1: Nodes are clustered in a single region of radius less than a wavelength}}\\
Consider a centralized control system in which the controller has downlink information for the users and the users have some uplink information for the controller and strict latency and reliability requirements are to be met. If all users are clustered in a single region of radius less than a wavelength, then all the channels to the controller are going to be highly correlated no matter where the controller is positioned.
Therefore, this scenario will result in having an overall failure rate greater than the tolerable rate of $10^{-9}$ -- if one node doesn't have a good channel to the controller, it is highly likely that other nodes also don't have good channels to the controller.
However, this is a worst case scenario where somehow all nodes land up in a tiny sphere and the \emph{only way} to combat this would be to transmit at a very high power.

\noindent\textbf{\emph{Case 2: Nodes are reasonably spread out in space}}\\
Again, the channel fades are going to be correlated.
However, does this correlation mean that the realizations end up being much worse than if they are independently distributed? Surprisingly, the answer is no.
Let us assume that we have the fade realization for the channel between the controller and a point $\vec{p}$ given by $h_p$ and we want to know the distribution of the channel fade between the controller and another point $\vec{q}$.
As these are jointly Gaussian, the channel fade between controller and
point $\vec{q}$ is given by $h_q|h_p \sim \mathcal{CN}(\rho_{||\vec{p}
  - \vec{q}||}h_p, \sigma^2 (1 - \rho^2_{||\vec{p} - \vec{q}||}))$
where $\rho_{||\vec{p} - \vec{q}||}$ is the correlation of the fade
distribution which depends on the distance between the two points
$||\vec{p} - \vec{q}||$ (as given by Eq.~\eqref{eq:bessel}), and
$\sigma^2$ is the unconditional variance of $h_q$ (also of $h_p$).

Given that we are in the realm of ``reasonably spread out in space'', we assume that $|\rho_{||\vec{p} - \vec{q}||}| < 0.2$ (the distance between nodes is at least $3$ times the wavelength as suggested by Fig.~\ref{fig:bessel}).
In such cases, the conditional variance remains largely unchanged i.e., $\sigma^2 (1 - \rho^2_{||\vec{p} - \vec{q}||}) > 0.96 \cdot \sigma^2 \approx \sigma^2$. However, the conditional mean can still change significantly from being zero mean to something else. If the channel between the controller and point $\vec{p}$ is deeply faded i.e., $h_p \approx 0$, then the conditional mean of $h_q|h_p$ given by $\rho_{||\vec{p} - \vec{q}||}h_p$ is close to $0$. In other words, \textbf{even if the channel $h_p$ is deeply faded, then the channel $h_q|h_p$ is distributed approximately as $\mathcal{CN}(0,\sigma^2)$ as if it was independently faded but with slightly lower nominal SNR.} If the channel between the controller and point $\vec{p}$ is not in deep fade, then the conditional mean shifts away from zero but again, the variance remains unchanged. In other words, \textbf{if the channel between the controller and point $\vec{p}$ is not in deep fade, then it essentially biases the conditional distribution at $\vec{q}$ towards a good channel but with a slightly smaller variance.}

These findings suggest that the effect of spatial correlation is not necessarily horrible.
In fact, a channel correlation of about $0.2$ can easily be modeled as a drop in nominal SNR of $0.05$dB.
Consequently, if the nodes are reasonably spread out, spatial correlation can provide fade realizations that are almost as good as or in some cases better than spatial independence.

\vspace{-10pt}
\section{Protocol Schemes}
\label{sec:protocol}
\begin{figure}[htbp]
\begin{center}
\includegraphics[width=.25\textwidth]{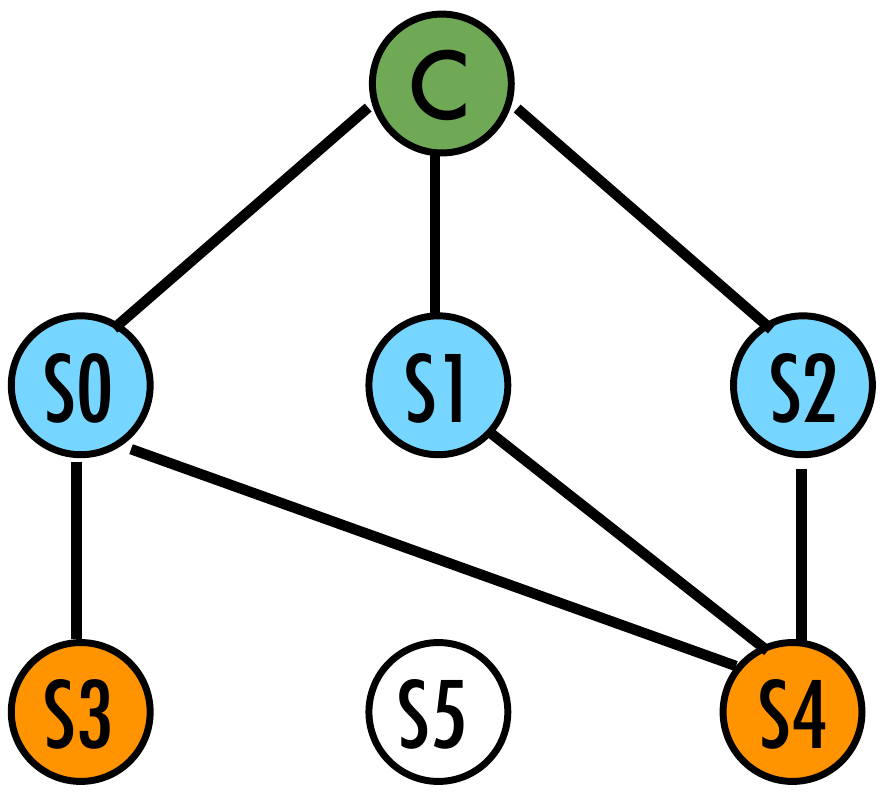}
\vspace{-10pt}
\caption{\small{An example of an instantaneously realized network topology for a network with one Controller and $6$ nodes -- $S0$ through $S5$. Nodes $S0$, $S1$ and $S2$ have good channels to the controller. Nodes $S3$ and $S4$ have good channels to the set of nodes that have good channels to the controller ($S0$, $S1$ and $S2$) and thus can get their messages potentially through this set. On the other hand, $S5$ has no good channels to any of these nodes and as a consequence does not get its downlink message from the controller, nor does the controller get its uplink message.}}
\label{fig:setup}
\end{center}
\vspace{-50pt}
\end{figure}

In this section, we very briefly describe the cooperative-communication-based protocols ``Occupy CoW'' and ``XOR-CoW''. We use these protocols to illustrate the impact of different channel characteristics by looking at the nominal SNR needed by these protocols to meet the latency and reliability requirements.
We consider a network with a central controller ($C$) that wishes to send and receive messages to and from each node in a set of $n$ nodes, denoted by the set $\mathcal{S}$ (see Fig.~\ref{fig:setup}). Distinct messages flow in a star topology from the central controller to individual nodes, and in the reverse direction from the nodes to the controller within a ``cycle'' of length $T$ (here $T$ is in the order of milliseconds). This cycle of communication must be completed successfully with a very small outage probability (on order of $10^{-9}$).

\subsection{Occupy CoW}
The Occupy CoW protocol uses multi-user diversity (introduced in~\cite{laneman2004cooperative}) to overcome bad fading events \cite{swamyTWC2017}. The main idea is to use a flooding strategy where a node broadcasts a packet with a set of nodes as the destinations. All nodes (destinations or not) try to listen to the message and nodes with good channels to the transmitter, decode the message and can act as relays.
All messages are scheduled such that each message gets a proportionate amount of time. 
The protocol can be programmed to have these initial transmissions be repeated a few times and interleaved together \emph{to guard against unmodeled error events} that are not just deep fade events. Let each message be transmitted by its source a total of $k_1$ number of times in the initial phases.

During these initial phases, all nodes are listening whenever they are not transmitting. Once all messages have been transmitted by their respective source $k_1$ times, the relaying phase begins. Again, in a scheduled manner, all nodes that have decoded the packet from the source ($S0, S1,$ and $S2$ in Fig.~\ref{fig:setup}) transmit it using a distributed-space-time code. Once again, the protocol can have the provision to enable multiple interleaved relaying slots per message, say $k_2$ of them, to guard against unmodeled error events isolated in time.
Destinations that did not get the packet directly will then try to decode the relayed transmissions ($S3$ and $S4$ in Fig.~\ref{fig:setup}). 
If there are any messages that have not reached their destination ($S5$ in Fig.~\ref{fig:setup}), then this causes a failure event.

We restate the basic error analysis from \cite{swamyTWC2017} for the above scheme \emph{under ideal conditions} such as quasi-static reciprocal channels and perfect knowledge of fade distribution, to illustrate the flavor of the protocol and bring attention to where the assumptions breaking down may be significant. We also assume that there are \emph{no unmodeled error events} and therefore $k_1 = k_2 = 1$ would be sufficient (no unmodeled error events mean we only need to combat deep fades). Consider a simple star information topology where there are $n$ nodes and one controller, with each node having a message for the controller and the controller having a distinct message for each of the nodes.
Denote the set of nodes with direct controller links by $\mathcal{A}$ and let the random variable associated with the cardinality of this set be denoted by $A$. Other nodes may connect to the controller through these nodes in a two-hop fashion.
Uplink transmissions are similar by reciprocity. Here, the transmission rate in each phase is $R_{w} = \frac{m n}{T/4}$. Therefore, the probability of a single link outage due to fading -- assuming a capacity-achieving code and a single-tap Rayleigh-faded channel is $p_{w} = 1 - \exp(-\frac{2^{R_{w}} - 1}{SNR})$. (Any repetitions would not help here because of the quasi-static assumption as well as the perfect nature of the assumed error-correcting code.) Then, the probability that $P(A = a)$ follows a Binomial distribution:
$$P(A = a) = \binom{n}{a}(1 - p_{w})^{a} p_{w}^{n-a}.$$

Under the \emph{ideal conditions that the channels do not change during a cycle and they are reciprocal}, the probability of cycle failure is the probability that at least one of the nodes in the set $\mathcal{S} \backslash \mathcal{A}$ does not connect to $\mathcal{A}$. Then we can write:
$$P(\text{fail} | A = a) = 1 - (1 - p_{w}^{a})^{n-a}.$$

Thus, the probability of cycle failure is given by:
\begin{align}
P(\text{fail}) &= \sum_{a = 0}^{n - 1} P(A = a) \cdot P(\text{fail} | A = a).
\label{eq:outage}
\end{align}

\subsection{XOR-CoW}
The XOR-CoW protocol~\cite{swamy2016cooperative} follows the same key ideas as the CoW-protocol described above, except that it also uses network coding inspired by works such as~\cite{katti2008xors,bagheri2011randomized}. This can be advantageous in the star information topology and increases efficiency by dividing the cycle length into only three phases. The downlink and uplink phase follow the same broadcast ideas as the Occupy CoW protocol. In the ``XOR'' phase, strong nodes that have both uplink and downlink messages for other nodes broadcast the XOR of the two messages, thus simultaneously serving as an uplink relay for the controller and as a downlink relay for other nodes.

Here, $R_{x} = \frac{m n}{T/3}$, which gives the probability of link failure $p_{x} = 1 - \exp(-\frac{2^{R_{x}} - 1}{SNR})$. This reduction of rate is the difference between the Occupy CoW and the XOR-CoW protocols, and  thus leads to the gains using this network coding based scheme. The expression for system outage is the same as Eq.~\eqref{eq:outage} with $p_{w}$ replaced by $p_{x}$.

\vspace{-10pt}
\section{Impact of Channel Dynamics and Uncertainty}
\label{sec:impact}
We have theoretically studied wireless channel dynamics and their temporal and spatial characteristics.
There are several viable candidates for enabling URLLC including~\cite{swamyTWC2017, jurdi2018variable, liu2018d2d}.
In this paper we described two candidate schemes Occupy CoW and XOR-CoW in Sec.~\ref{sec:protocol} and analyzed their performance under \emph{ideal conditions}. However, to make any scheme practical, we need to understand the effects of real world imperfections. Consequently, we need to understand the effects of different \emph{unmodeled events} that could potentially cause severe degradation. This is crucial as \textbf{wireless systems supporting URLLC applications must build in robustness}.
In order to build a robust communication system, we must ask the following questions: a) what events may cause errors, b) how can we model the effects of these errors on the communication system, and c) what avenues does the communication system have to protect against these events?

Let us start by answering the first two questions: what events may cause errors and how can we bound their effects on the communication system? This will allow us to create an appropriately sized and shaped ``uncertainty ball'' around the nominal wireless model. The resulting uncertainty-bounding parameters are summarized in Table~\ref{table:allKnobs}.
\begin{itemize}[
    \setlength{\IEEElabelindent}{\dimexpr-\labelwidth+8pt}
]
\item The dominant cause of error is deep fading. The frequency of occurrence of deep fades can be modeled using the nominal fading distribution. However, how much do we depend on the accurate knowledge of the fading distribution? What happens if fades actually come from a different distribution? We account for this using an additional probability of error $\leq p_{off}$ that is independent across links and find that its effect is small (i.e.~a little bit of added SNR compensates for it) for medium - large network sizes (Sec.~\ref{sec:offset}).

\item Imperfections in different quantities of interest could potentially cause errors as well. What are some of the quantities/measures that can be imperfect? An obvious and significant one is clock imperfection -- mis-synchronized clocks at transmitters and/or receivers can lead to decoding errors which can effectively destroy an entire slot. Channel estimation errors also can have this impact. Another similar cause of errors is abrupt channel changes during a packet.  (e.g. a transmitter moving in such a way that it transitions to being shadowed to scatterers or loses/gains line-of-sight to the receiver.) This can also lead to incorrect decoding of the message. We bound these kinds of errors on an independent \emph{per-slot per transmitter basis using $p_{c}$ while ensuring that when there are more nodes simultaneously transmitting during a slot,  the chance of encountering these kinds of unmodeled errors grows} (Sec.~\ref{sec:packetchange}).

\item Another potential cause of errors is interference from other devices in the vicinity. This could come from a network nearby in which a node accidentally transmits at a very high power -- thereby causing a burst of interference throughout our network. A failure of the error-correcting code due to an unlucky realization of additive noise is similar, and so is motion of the receiver that causes it to transition into a shadow relative to important scatterers.  
We bound these kinds of events as well on an independent \emph{per-slot basis using $p_{g}$, but in a way that does not compound with the number of simultaneous transmitters}.
\end{itemize}

We now answer the last question: what avenues does the communication system have to protect against these events? For errors such as the decoding error caused due to clock mis-synchronization causing packet collisions, these can be combated only by doing retransmissions. It is not that the channel between the transmitters and receiver was faded but rather there was an \emph{uncontrollable error} that caused the transmission to fail. Therefore, \textbf{retransmissions or time margin} is the way to combat these events that are not about SNR. What about the errors caused due to fading? As mentioned earlier, spatial correlation could potentially lead to slightly worse channel realizations than an independent realization. The only way to improve the channel itself is by increasing the transmit power to get better nominal SNRs. Therefore, to combat channel-fade-related events, we use \textbf{SNR margins}. Our model based on the behavior of multipath establishes that the multipath fades only change slowly across time relative to the cycle time, and so all temporal correlations can also be treated the same way, using a small SNR margin while assuming that good channels stay static during the cycle.

With all wireless effects accounted for in either the nominal model or the uncertainty bounds, we can have some confidence that a system which performs well in theory will indeed be ultrareliable in practice. In the subsequent subsections, we look at these effects more closely.

\begin{table}[htb]
\small
\begin{center}
    \begin{tabular}{ | c | c | p{12cm} |}
    \hline
    Parameter & Range & \hspace{4cm} Unmodeled events captured \\ \hline
    $p_{off}$ & $0 - 0.1$ & Imperfections in channel fade modeling. Spatial-correlation-based degradation can also be captured through this term. \\ \hline
    $p_{c}$ & $0 - 10^{-2}$ & Errors due to clock mis-synchronizations or channels changing rapidly during a single packet. These errors compound with increasing number of simultaneous transmitters. \\ \hline
    $p_{g}$ & $0 - 10^{-2}$ & Global errors that are due to burst-interference like events. These errors do not compound with increasing number of simultaneous transmitters. They fundamentally exist at the receivers. \\ \hline
    \end{tabular}
\end{center}
\vspace{-20pt}
\caption{\small{Uncertainties captured and the parameters associated with them. Except for $p_{off}$, all of these are essentially independent from time-slot to time-slot if we assume that the communication scheme interleaves repetitions. Some of these might implicitly depend on the length of the time-slots (growing with time-slot length), but this dependence is suppressed here.}}
\label{table:allKnobs}
\vspace{-40pt}
\end{table}



\subsection{Effects of uncertainty in channel fade distribution}
\label{sec:offset}

Bad multipath fades have been modeled to be the dominant cause of potential failures. We assumed that the channel fades themselves come from a Rayleigh distribution. Given the extremely low error probabilities we are targeting in a wireless setting, it is natural to question if we can really trust the fading distribution down to $10^{-9}$? What happens if there are unmodeled events (e.g.~the exact geometry of the scatterers in the environment) that cause bad fades to occur more often than we had modeled. (We do not have to worry about events like line-of-sight paths that make bad fades occur less frequently.)
To capture this, we introduce \emph{an extra probability of failure at each link}, $p_{off}$, on top of the probability of error due to nominal fading, $p_{w}$. In this model, the link failure comprises of two parts: one coming from the nominal fading distribution and the other from local modeling error, $p_{off}$, is $p_{link}=p_{w}+p_{off}$. This is a local error model where each link gets affected independently -- i.e., unmodeled errors themselves are not correlated. Because this bound $p_{off}$ attaches to the individual link fades, we do not assume that it is realized independently across different time-slots in which that same link is active.

Consider the Occupy CoW scheme as described in Sec.~\ref{sec:protocol} with $n$ nodes, each sending messages of $m$ bits and total cycle time of $T$.
Nominally, links are modeled as failing if the fade was deep enough i.e., the instantaneous capacity given by $C_{\text{inst}} = W\log(1 + |h|^2SNR)$ is less than the rate of transmission. The probability of a bad link under this perfect Rayleigh channel fade distribution model is $p_{w} = 1 - \exp(-\frac{2^{R_{w}} - 1}{SNR})$ where $R_{w} = \frac{m n}{T/4}$.

\begin{figure}
\begin{center}
\begin{subfigure}[b]{0.48\textwidth}
\centering
\includegraphics[width=0.9\textwidth]{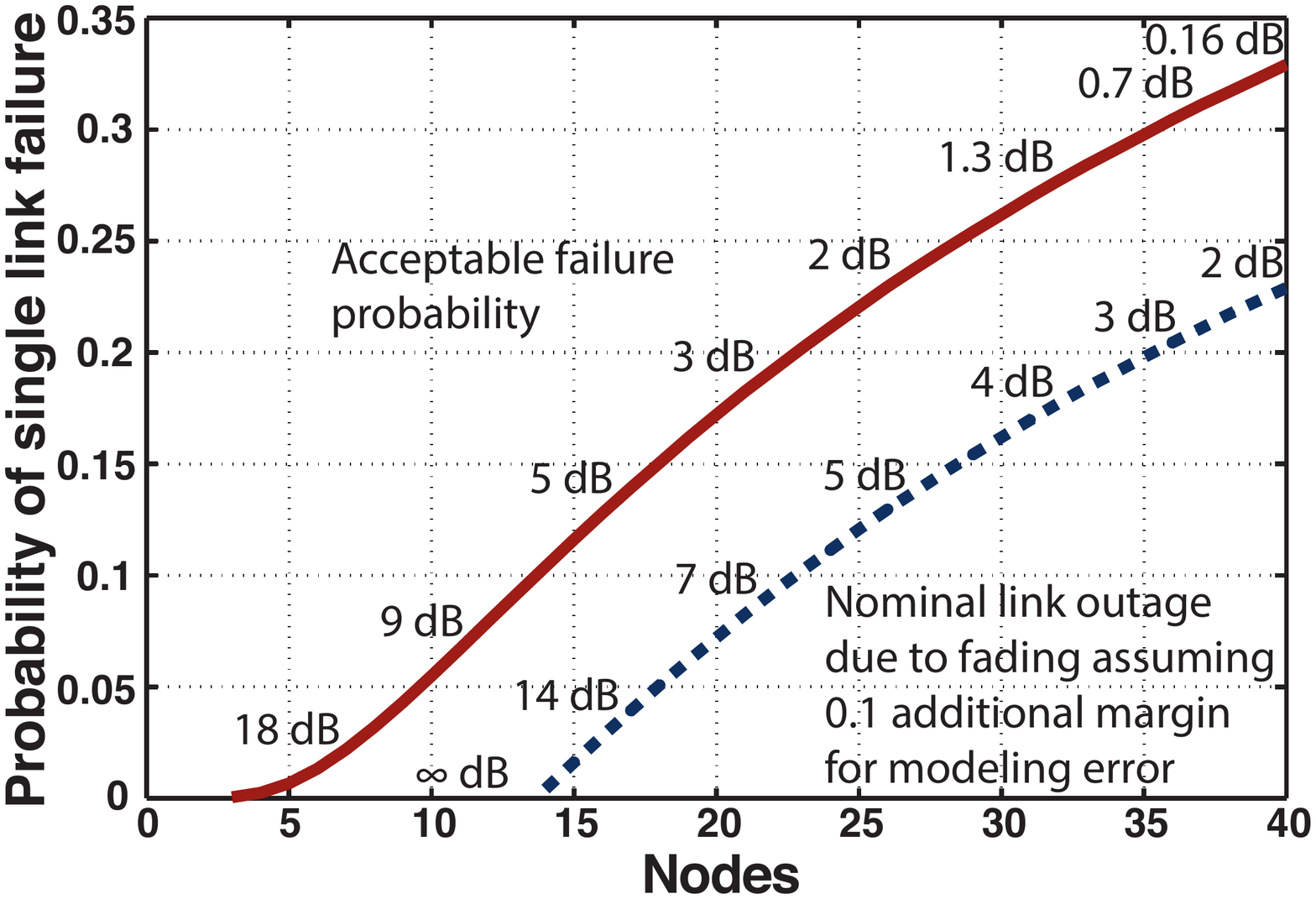}
\caption{The probability of link failure that can be tolerated for Occupy CoW as a function of the number of nodes. The lower curve is $0.1$ below and the SNR numbers represent the nominal SNR required to hit that particular link failure probability for Rayleigh fading.}
\label{fig:singlelink}
\end{subfigure}
~
\begin{subfigure}[b]{0.48\textwidth}
\centering
\includegraphics[width=0.9\textwidth]{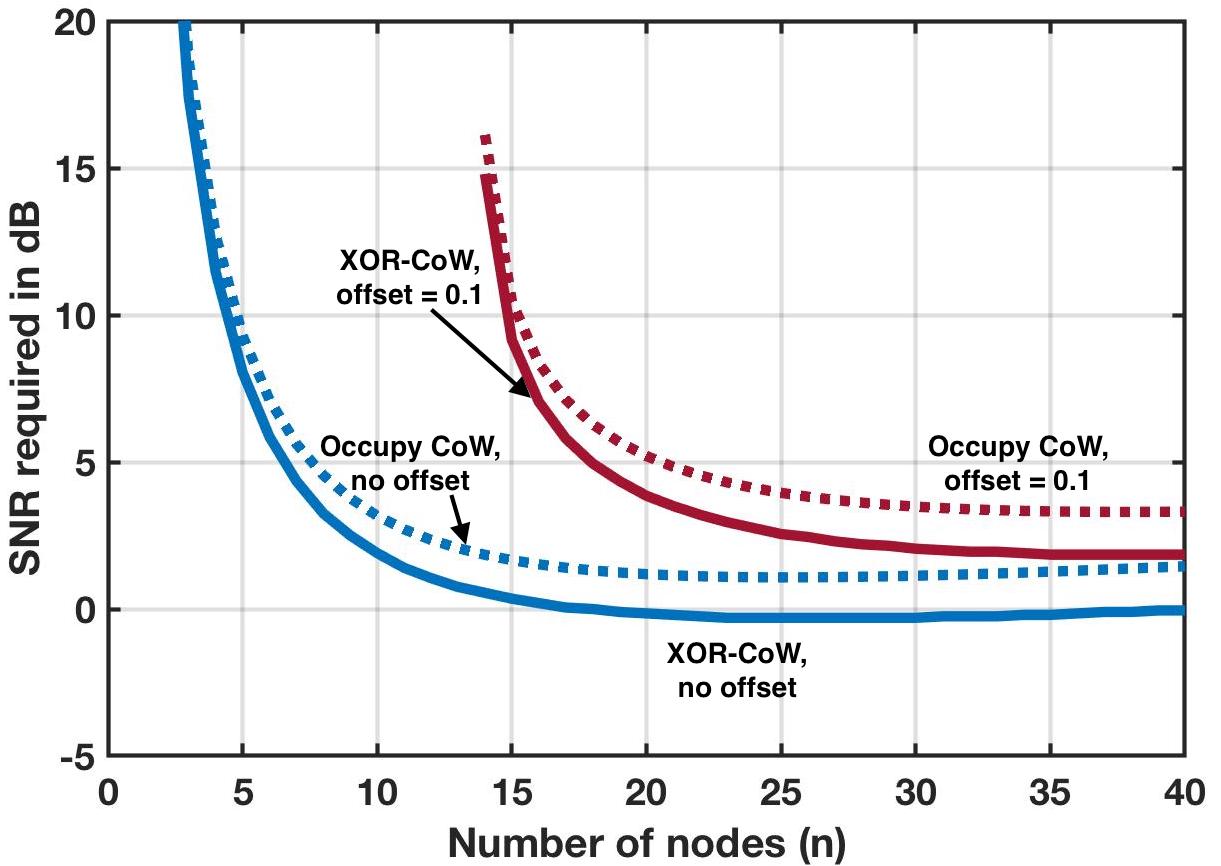}
\caption{SNR paid to achieve performance robustly in the face of uncertainty. The effects are similar for XOR-CoW. Below 14 nodes, it is not possible to be robust to the $p_{off} = 0.1$ of unmodeled uncertainty specified here.}
\label{fig:snrpernalty}
\end{subfigure}
\end{center}
\caption{\small{Effects of unmodeled errors on the performance of Occupy CoW and XOR-CoW. We assume the availability of a $20$MHz bandwidth channel and every message is of $160$ bits long.}}
\vspace{-40pt}
\end{figure}

We can look at the \emph{maximum} value of $p_{link}$ that can be tolerated for different number of nodes while keeping the overall probability of failure constant at $10^{-9}$ in the top curve in Fig.~\ref{fig:singlelink}. This tells us that if $p_{link}$ is greater than the error in modeling error $p_{off}$, then increasing $SNR$ to make $p_w$ smaller would be able to digest the modeling error. We see that if $p_{off} = 0.1$, then shifting the maximum tolerable $p_{link}$ down by $p_{off}$ will give us the maximum $p_{w}$ that can be tolerated which ultimately translates into an increase in the minimum SNR required. We note that for larger tolerable $p_{link}$, the SNR penalty is smaller (compare the SNR penalty for network size $30$ and $15$). 
In addition to incurring an SNR penalty, for smaller network sizes with maximum tolerable $p_{link} < p_{off}$ i.e., $N \leq 13$, there exists no SNR that can robustly support these requirements.

XOR-CoW has similar response to channel fading distribution uncertainty (Fig.~\ref{fig:snrpernalty}).
We conclude that unmodeled local errors such as not having perfect knowledge of fading distributions do not cause heavy degradation in the performance of schemes that rely on the availability of independently faded links. In fact, channel-correlation-induced extra link failure can be captured in $p_{off}$ as ultimately the effects of both are the same -- reduction in nominal SNR while `preserving' essential independence across space.

\vspace{-10pt}
\subsection{Effects of channel changes within a packet}
\label{sec:packetchange}

\begin{figure}
\begin{subfigure}[b]{0.48\textwidth}
\centering
\includegraphics[width = \textwidth]{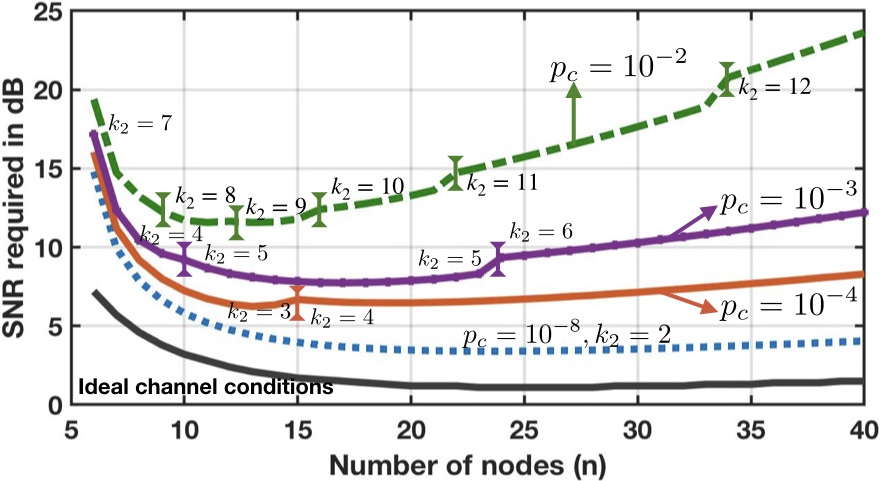}
\caption{Performance of Occupy CoW when the channel changes within a packet (transmitter centric) causing decoding errors. Here $p_g = 0$ and $p_{off} = 0.01$.}
\label{fig:midpacket}
\end{subfigure}
~
\begin{subfigure}[b]{0.48\textwidth}
\centering
\includegraphics[width = \textwidth]{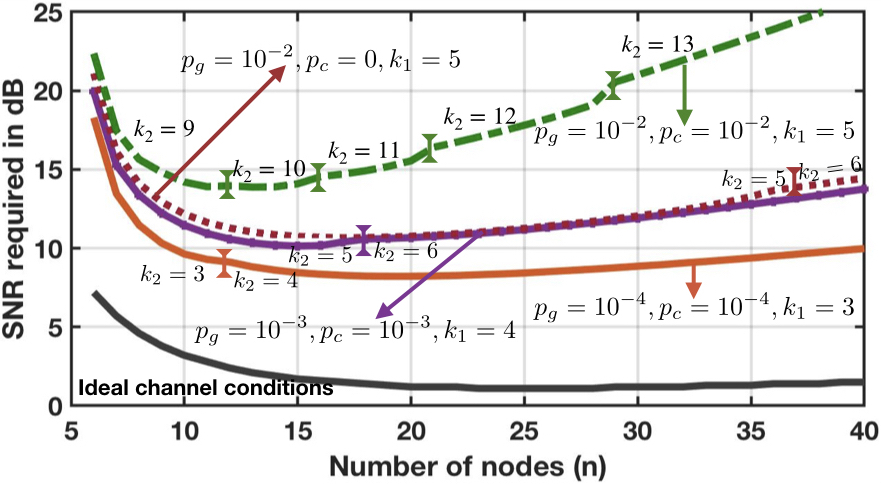}
\caption{Performance of Occupy CoW when receiver-center, transmitter-centric and fading uncertainty errors occur. Here $p_{off} = 0.01$.}
\label{fig:allKnobs}
\end{subfigure}
\caption{\small{Effects of unmodeled per-slot errors modeled by $p_g$ (non-cumulating) and $p_c$ (cumulating). The curves are annotated with the $k_1$ and $k_2$ needed to achieve the performance at that minimum nominal SNR. The exact expressions used to obtain these results can be found in the appendix~\cite{URLLCAppendix}.}}
\vspace{-40pt}
\end{figure}

In Sec.~\ref{sec:temporalpacket}, we studied the temporal characteristics of wireless channels within the packet duration. We saw that good channels tend to remain good for short packets if multipath is the only effect causing fades. 
However, a rapid change in the channel coefficient within a single packet, say due to crossing into a shadow of an obstacle relative to many scatterers, when the channel is good could lead to decoding errors.
What would be the effect of these errors?

Our approach is to bound the probability that a packet in a time-slot is corrupted by some maximum probability of such corruption. There are clearly two qualitatively different kinds of corruption that we need to watch out for. One is where the corruption happens ``at the transmitter'' --- for example, if the transmitter had moved in such a way that its channel rapidly transitioned into or out of a shadow relative to say a line-of-sight path. When multiple nodes are transmitting simultaneously, the receiver is decoding using the combined channel which depends on the DSTC and the individual channel realizations. If any of the channels change during the transmission (causing a corrupt packet to effectively be sent), it could potentially lead to decoding errors. In fact, the more nodes that transmit simultaneously, the more likely one of the channels would change mid-packet causing a decoding error. We bound this error using $p_c$ which has a \emph{cumulating effect} when there are more number of nodes transmitting during a single message slot. Essentially, if $r$ nodes are transmitting in a single slot then the probability of slot success is $(1 - p_c)^r$ However, we assume that these are independent from one message slot to the next.

The other qualitatively different kind of corruption happens ``at the receiver'' --- for example, if the receiver is the one that moves suddenly into or out of a shadow relative to a particularly important scatterer. These errors do not cumulate with multiple simultaneous transmissions. We bound this error using $p_g$ with a probability of slot success being $(1-p_g)$. This is also modeled as being independent from one message slot to the next.

The advantage of these kind of unmodeled uncertainty bounds is that they can encompass many different physical sources of imperfection. For example, channel estimation errors at the receiver can contribute to both $p_c$ and $p_g$ depending on how the pilots and preambles are transmitted. Synchronization errors are clearly a part of $p_c$. Interference bursts and imperfections of the error-correcting codes are just as clearly a part of $p_g$. For all of these, the important thing is that these phenomena (just like the feared rapid transition into a shadow) are finely localized in time.

Because they are finely localized in time, it is imperative to take advantage of time margins here to be robust to them. These unmodeled events are considered\footnote{Our goal is to make a wireless system ultrareliable to the impairments for which there is hope of being robust to. If a node were to turn into a persistent jammer, we cannot protect against that, the same as not being robust to placing one of the nodes in a Faraday cage. Those kinds of impairments are both unmodeled and irrelevant for wireless ultrareliability.} as being independent across slots (this is what the assumption of interleaved repetitions justifies at the level of each message), so there is a time diversity of sorts vis-a-vis unmodeled corruptions. This is unlike the traditional notion of time diversity where fading channels change from one slot to the other. Here, the channel quality (being a good/bad channel) remains the same across slots but these other errors happen independently across those same slots. We see that to combat these events, we need to have \emph{multiple relaying slots} for each message, i.e.,  $k_1 > 1$ and $k_2 > 1$, where $k_1$ is the number of times a message is transmitted initially and $k_2$ is the number of times the message is transmitted in the relaying phases. These $k_1,k_2$ are not the same as hops in a multihop protocol.


We illustrate this in Fig.~\ref{fig:midpacket} where we see that when $p_c$ is super low $\approx 10^{-8}$, the effect is almost negligible. However, we see a very interesting phenomenon for mid-high bounds $10^{-4} - 10^{-2}$. 
We optimize over different values for $k_2$ and pick the one that minimizes the nominal SNR. The curves associated with $p_c = 10^{-2}$ and $p_c = 10^{-4}$ are annotated with this optimal number of $k_2$.
\vspace{-10pt}
\subsection{Effect of channel change during a cycle}
We studied to what extent channels change within a cycle and redefined the notion of ``coherence time'' in Sec.~\ref{sec:temporalcycle}. In this section, we address the following question: if all available relays were to be employed, what are the effects of significant channel changes during a cycle (henceforth referred to as non-quasi-static channels) if channels remain completely static during a single packet transmission (so the effects seen in Sec.~\ref{sec:packetchange} do not occur)?
As mentioned earlier, the easiest way to account for this is to fold these rare events into the $p_{off}$ term earlier.
However, it is possible to analyze this even more conservatively. Here, we briefly examine the performance when channels refresh at phase boundaries of the protocol: for eg., it might change between the downlink and uplink phases for the Occupy CoW protocol. This effectively translates any changes during a cycle/phase into an easier to analyze effect.

Such non-quasi-static channels introduce more randomness into the system. In the Occupy CoW protocol, this extra randomness might give some nodes two chances to directly establish a link to the controller, before and after a mid-cycle channel change. This means that downlink-only or uplink-only performance of a protocol can improve due to the extra diversity introduced by a channel change.
However, this breaks the assumption of reciprocity and consequently, the combined performance of uplink \& downlink take a small hit. In the quasi-static case, a path that worked for two-hop downlink to a node was guaranteed to work for two-hop uplink for the same node.
In the presence of changing channel fades, this is no longer true.
Each node must potentially find two independent paths to the controller --- one for uplink \& one for downlink.

The hit for the XOR-CoW protocol is more pronounced. The performance hit is due to the decoupling between uplink and downlink --- this can lead to a smaller set of nodes that have both uplink and downlink information for any given node --- and thus a smaller set of nodes that can help anyone who does not have a direct link to the controller. The degradation in performance is captured in Fig.~\ref{fig:qs}. 
The key takeaway is that this entire effect is small even in the worst case, and only costs an SNR margin of a little over a single dB.

\vspace{-10pt}

\subsection{Combined effect of all error events}
Until now, we have analyzed the impact of different kinds of events and phenomenon \emph{individually}. 
It is important to put together all these events and analyze the combined effects to understand how much it costs to get the robustness we need by budgeting the SNR and time margins appropriately. We capture this in Fig.~\ref{fig:allKnobs} where we account for the following events: a) deep fade causing links to be bad captured by the nominal model for $p_w$, b) bounded uncertainty in fading model $p_{off} = 0.01$, c) global per-slot bounded badness such as interference, error-correcting-code failures, or receiver shadow transitions that does not cumulate with the number of transmitters $p_g$ (different values explored), and d) bounded per-transmitter badness due to say mis-synchronized clocks, channel estimation errors, or transmitters transitioning into shadows $p_c$ (different values explored) which cumulates with the number of transmitters. The exact formula used to obtain the curves can be found in the appendix~\cite{URLLCAppendix}.

We immediately notice: the number of retransmissions required in the initial phase $k_1$ is primarily dependent on $p_g$ as that is the main unmodeled event to guard against in the initial phase as there are no simultaneous transmissions. 
As $p_g$ and $p_c$ increase, we see increases in the number $k_2$ of retransmissions in the relaying phase. The increased retransmissions induce a need for higher raw spectral efficiency which drives up the SNR required. In fact, if we consider $p_g = p_c = 10^{-2}$ essentially uncontrollable unmodeled events occurring $1 \%$ of the time such as shadowing transitions and budget an extra $3$dB for finite-blocklength codes, we see that we need to roughly operate in the regime of $15$dB to $20$dB nominal\footnote{As mentioned earlier, if we wanted to be extra paranoid and consider fading environments that might just have two evenly balanced scatterers, then all the SNR requirements would further increase by a little over $10$dB.} SNR to be robust to most realistic error events, whereas under ideal channel conditions, we needed to be around $3$dB. Most of this is due to the increased bitrate needed to support the repeated transmission of the small packets.

\vspace{-10pt}

\subsection{Why the nominal model matters}
In this last subsection, we argue why it is important to have picked a nominal model that took some care to understand how the spatial distribution of fades gives rise to reliable multiuser diversity. However, if we had failed to model the effects of spatial correlation carefully or if we had not considered unmodeled events and budgeted for them through time-margins and instead took a pessimistic approach, how would the penalties look like? We briefly consider these two scenarios here and point the reader to~\cite{swamy2016robustness} for a detailed treatment.

\subsubsection{Effect of spatial correlation of channels}
\label{sec:spatialcorrel}
\begin{figure}
\begin{center}
\begin{subfigure}[b]{0.31\textwidth}
\centering
\includegraphics[width = \textwidth]{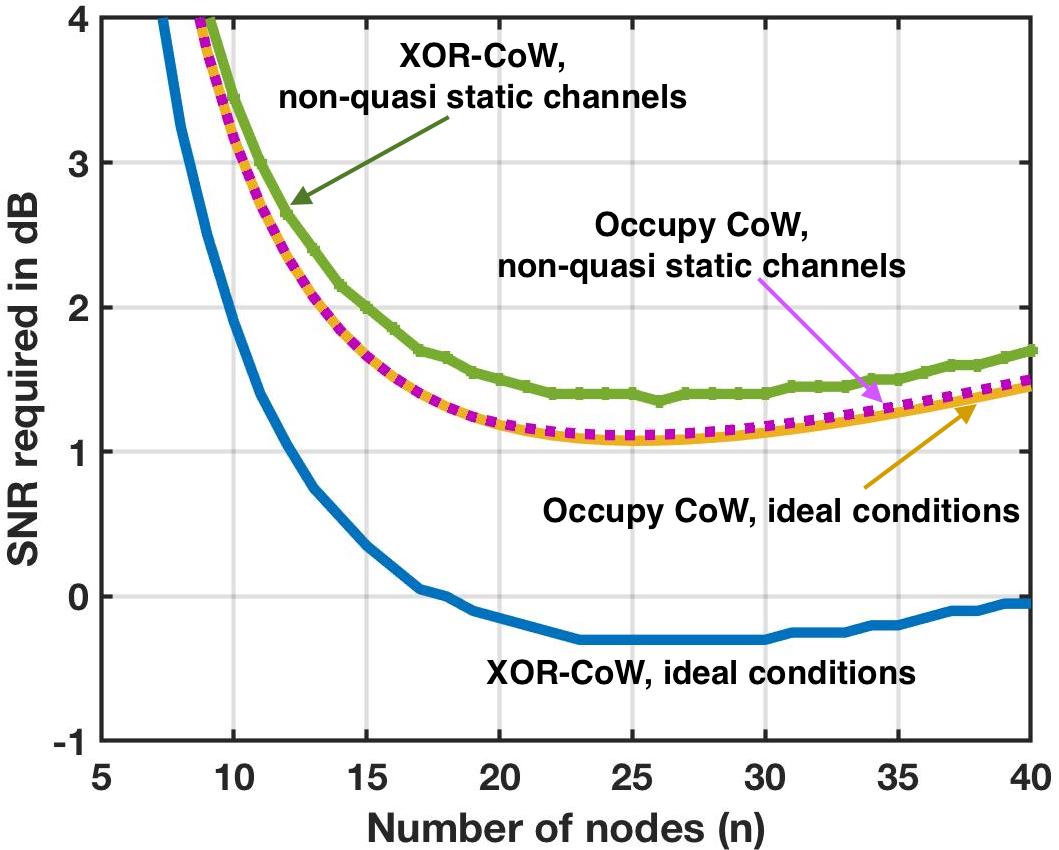}
\caption{Performance of Occupy CoW and XOR-CoW protocols when the channel changes during a cycle but not within a packet. New fades are realized which breaks reciprocity.}
\label{fig:qs}
\end{subfigure}
~
\begin{subfigure}[b]{0.31\textwidth}
\centering
\includegraphics[width = \textwidth]{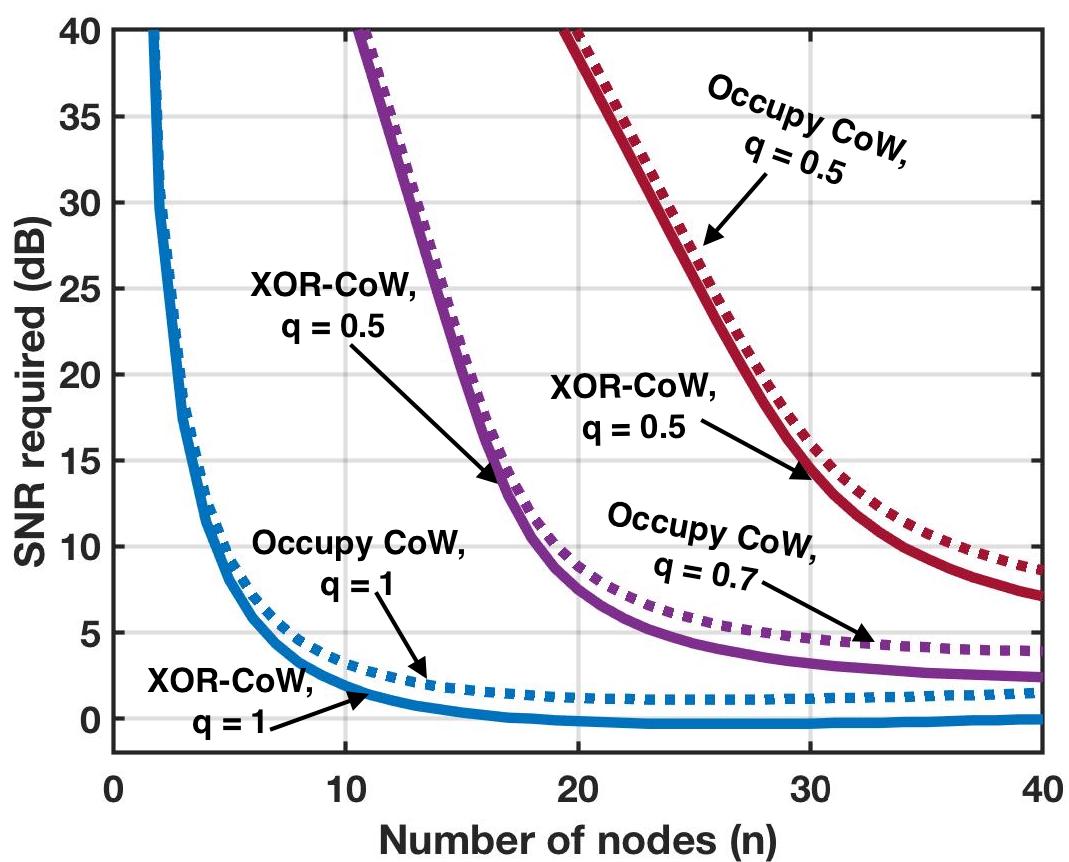}
\caption{Performance of Occupy CoW and XOR-CoW protocols using a pessimistic spatial correlation model. $q$ represents the probability of an independent fade on a channel.}
\label{fig:independence2}
\end{subfigure}
~
\begin{subfigure}[b]{0.31\textwidth}
\centering
\includegraphics[width = \textwidth]{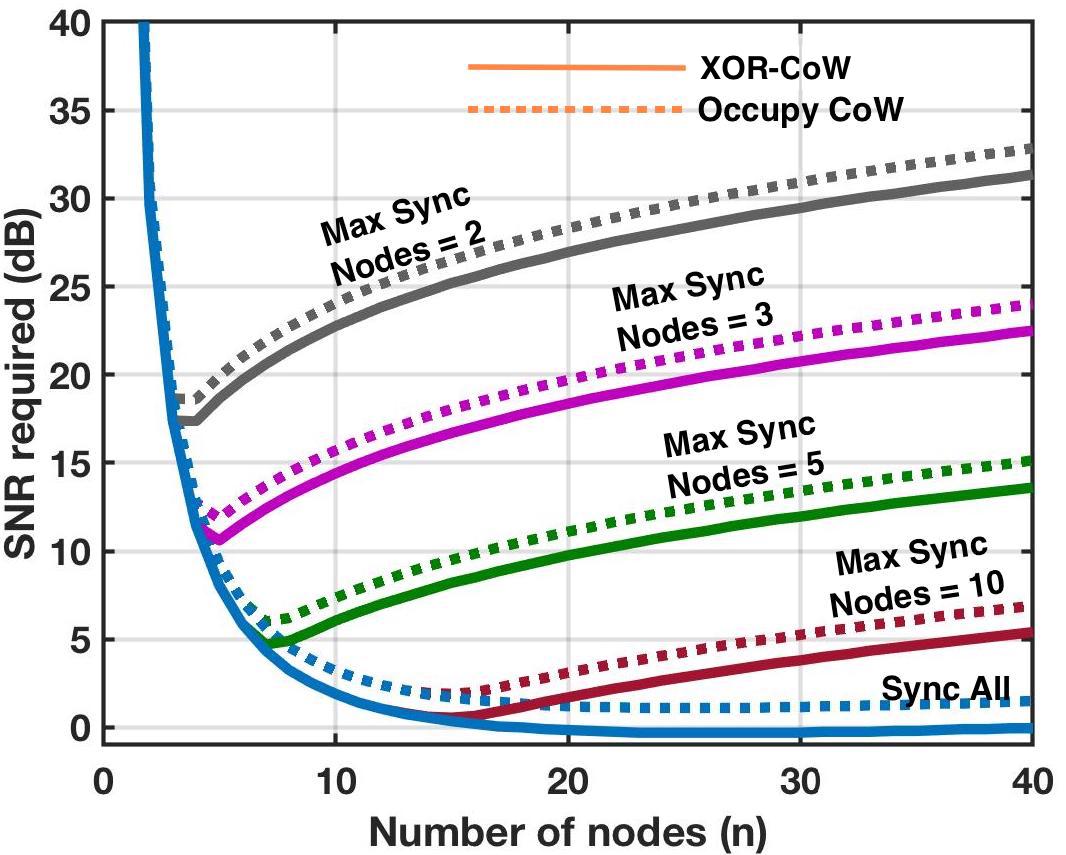}
\caption{\small{Performance of Occupy CoW and XOR-CoW protocols with a cap on the number of nodes that can transmit simultaneously. Here, $k_1 = k_2 = 1$ and there are no repetitions.}}
\label{fig:independence1}
\end{subfigure}
\end{center}
\vspace{-30pt}
\caption{\small{Effects of non-quasi-static channel realizations and pessimistic spatial correlation modeling}}
\vspace{-40pt}
\end{figure}

We studied spatial correlation of channels in Sec.~\ref{sec:spatial} and saw that the channels are indeed correlated, but since we can assume that nodes are more than $2\lambda$ apart, this correlation actually only leads to a small degradation in nominal SNR -- of about $0.05$dB. So a nominal model of independence is justified.
Let us now construct a pessimistic model about spatial correlation and understand the effects of such a model.
Every new channel has a probability $q$ of coming from an independent fading realization and with probability $1-q$ the channel is identical to a channel that has already been realized, so we get no diversity. This might sound reasonable, but Fig.~\ref{fig:independence2} demonstrates how this affects cooperative-communication-based URLLC protocols.
The SNR curves decay as the number of nodes increases but a low probability of independence has a severe impact. Especially for smaller networks, around $20$ nodes, the SNR penalty is about $40$dB. However, from Sec.~\ref{sec:spatial} we know that when the nodes are sufficiently separated, then actual multipath channel realizations, although correlated are not too much worse than completely independent realizations.

\subsubsection{Effect of synchronization impediments}
Timing and frequency synchronization pose the biggest hurdles in making practical cooperative communication systems as the performance of most synchronization algorithms degrades with more relays~\cite{li2009timing}.
We have so far bounded this effect using $p_{c}$ which corresponds to unmodeled error terms that cumulate with simultaneous transmitters. Because of this cumulative behavior, we think that a good wireless communication system should try to seek a kind of ``sparsity'' to be robust, analogous to what the analysis of wideband channels suggests for traditional communication~\cite{verdu2002spectral}. It is safer and simpler to avoid too many simultaneous transmissions. (In~\cite{swamy2018predicting}, we provide ways to leverage data-driven learning to greatly reduce the need for many simultaneous transmissions, but omit those here due to space constraints.)
However, implementation constraints might not behave in the cumulative manner that we assume for $p_c$. Therefore, we analyze the effect of restricting the total number of simultaneous transmissions to some maximum number dictated by the synchronization protocol.


Fig.~\ref{fig:independence1} shows the significant impact on performance when the maximum number of simultaneous transmitters for each message is capped.
We see the SNR increase with increasing number of nodes because the additional nodes in the system stop being useful as new relays which leads to the increase in nominal SNR required. Again, we may be able to combat this through time margins by having multiple relaying slots with a smaller number of transmitters per slot.

\vspace{-10pt}
\section{Conclusions}
\label{sec:conclusion}
In this paper, we examine channel dynamics in the URLLC context and
proposed a robust-control inspired nominal plus uncertainty-bound
perspective to wireless environment modeling. For the nominal model,
we refined the standard Jakes' model-based view of Rayleigh fading and
established that as long as the wireless nodes are always separated by
a few wavelengths, the assumption of spatial independence essentially
holds. Furthermore, we showed that although the traditional view of
the fading process as being strictly bandlimited is false, the
channel variation within a single short packet is very small once we
condition on the channel being good to begin with. Across the entire
low-latency cycle, the variation is more substantial. This does not
greatly impact the effectiveness of relaying-based strategies however,
as long at they are able to use multiple potential relays per message.

With the fear of unfortunate spatial correlations exorcised, we
bounded the rest of the uncertainty using three terms, each of which
we believe encompassed events that would be independent across time-slots: (a) a bound on the probability of unmodeled receiver-centric
temporary outages of time-slots; (b) a bound on the probability
of unmodeled transmitter-centric temporary corruptions that would
compound if there were multiple simultaneous transmissions in a single
time-slot; and (c) a bound on the additional unmodeled probability of
fading for a given channel. We argued that between
all of these bounds, interference, error-correcting code issues,
synchronization issues, shadowing transitions, and other propagation
effects are all covered.

To be robust against the first two, a communication scheme
has to have time margin by repeating the same message using
interleaved slots\footnote{By identical logic, we could also have added a fourth
  uncertainty bound for unmodeled corruption of a frequency slot that
  might span many time slots. The same schemes would
  work except they would also need to hop between frequencies
  as they hopped into time slots. The required number of such hops would not
  have to increase if the uncertainty bound for frequency corruption
  was not bigger than the bound for time-slot corruption since what
  the hopping is seeking is an independent chance to experience
  something closer to the nominal model. Furthermore, as long as the
  URLLC system insisted that only one of its own messages was being
  transmitted at any given time, there is clearly no additional
  counterpart of $p_{c}$ for compounding errors for per-channel
  unmodeled dynamics. However, if multiple messages were
  transmitted in overlapping times but in different channels, then
  such a term would need to exist --- although it would be small if
  the frequency slots were very well separated.}. For
the last, an increased SNR margin is required. The use of repetitions
also increases the SNR required since messages must be successfully
communicated using shorter slots and thus higher spectral efficiency.
The combination of explicit modeling of known effects and bounding
channel uncertainty in a way that captures the ``shape'' of wireless
protocols  allows us to have confidence in ultra-reliability.

The fact that URLLC systems can be made robust to unmodeled uncertainties at the $10^{-2}$ to $10^{-3}$ level means that such uncertainties can be tracked and learned through data-driven self-monitoring of a wireless system.
After all, they will manifest as small anomalies where a packet that
didn't succeed in one repetition does in another, despite nominally
facing the same channels, multiple times per second. By contrast, a
$10^{-9}$ event will not be seen with any reasonable frequency to
support learning.
Consequently, learning actually might have an important role to play in allowing systems to reduce their spectral and time footprint while maintaining reliability, if they can assume that these unmodeled uncertainties are not going to change abruptly.

\section*{Acknowledgements}
\noindent Thanks to Adam Wolisz, Ajith Amerasekera, Venkat Anantharam and Meryem Simsek for useful
discussions. We thank BWRC students, staff, faculty and sponsors, NSF grants AST-1444078, CNS-1321155, and ECCS-1343398
and Microsoft for a Microsoft Research Dissertation Grant.

\bibliographystyle{IEEEtran}
 {\small{
 \bibliography{IEEEabrv,Cow}}

\end{document}